\documentclass{article}

\usepackage{lineno,hyperref}
\modulolinenumbers[5]
\usepackage{arxiv}
\usepackage{amsmath,amsthm}
\usepackage[utf8]{inputenc} 
\usepackage[T1]{fontenc}    
\usepackage{hyperref}       
\usepackage{url}            
\usepackage{booktabs}       
\usepackage{amsfonts}       
\usepackage{nicefrac}       
\usepackage{microtype}      
\usepackage{lipsum}		
\usepackage{graphicx}
\usepackage{natbib}
\usepackage{doi}
\usepackage{algorithmic}
\usepackage{textcomp}
\usepackage[ruled,vlined]{algorithm2e}
\usepackage{caption}
\usepackage{subcaption}
\usepackage{listings}
\usepackage{multirow}
\usepackage{multicol}
\usepackage{pgfplotstable}
\usepackage{pgfplots}
\pgfplotsset{
	compat=1.16,
	/pgf/number format/1000 sep=
}
\theoremstyle{definition}
\newtheorem{definition}{Definition}

\title{A Game-Theoretic Approachfor AI-based Botnet Attack Defence}

\author{ {\includegraphics[scale=0.06]{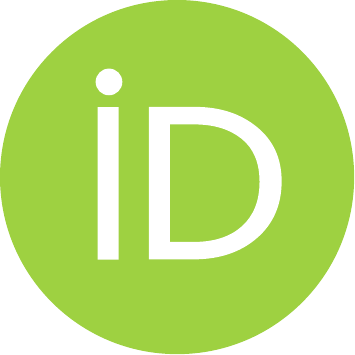}\hspace{1mm}Hooman Alavizadeh} \\
	Cybersecurity Lab\\
	Comp Sci/Info Tech, Massey University\\
	Auckland, NEW ZEALAND \\
	\And
	{\includegraphics[scale=0.06]{orcid.pdf}\hspace{1mm}Julian  Jang-Jaccard} \\
	Cybersecurity Lab\\
	Comp Sci/Info Tech, Massey University\\
	Auckland, NEW ZEALAND \\
    \texttt{j.jang-jaccard@massey.ac.nz} \\
    \And
	{\includegraphics[scale=0.06]{orcid.pdf}\hspace{1mm}Tansu Alpcan} \\
	Department of Electrical and Electronic Engineering\\
	University of Melbourne\\
	Parkville, AUSTRALIA \\
    \And
	{\includegraphics[scale=0.06]{orcid.pdf}\hspace{1mm}Seyit Camtepe} \\
	Data61 CSIRO\\
	AUSTRALIA \\
}

\renewcommand{\shorttitle}{\textit{arXiv} Template}

\hypersetup{
pdftitle={A template for the arxiv style},
pdfsubject={q-bio.NC, q-bio.QM},
pdfauthor={David S.~Hippocampus, Elias D.~Striatum},
pdfkeywords={First keyword, Second keyword, More},
}

\begin{document}
\maketitle

\begin{abstract}
	 The new generation of botnets leverages Artificial Intelligent (AI) techniques to conceal the identity of botmasters and the attack intention to avoid detection. Unfortunately, there has not been an existing assessment tool capable of evaluating the effectiveness of existing defense strategies against this kind of AI-based botnet attack. In this paper, we propose a sequential game theory model that is capable to analyse the details of the potential strategies botnet attackers and defenders could use to reach Nash Equilibrium (NE). The utility function is computed under the assumption when the attacker launches the maximum number of DDoS attacks (i.e., to maximize the probability that the target system become unavailable) with the minimum attack cost while the defender utilises the maximum number of defense strategies (i.e., to maximize the probability that the victim’s system recovers) with the minimum defense cost.  We conduct a numerical analysis based on a various number of defense strategies involved on differnet (simulated) cloud-band sizes in relation to different attack success rate values. Our experimental results confirm that the success of defense highly depends on the number of defense strategies used according to careful evaluation of attack rates.

\end{abstract}


\section{Introduction}
A strong cyber defense system should be able to detect, monitor, and promptly leverage defence mechanisms to the cyber threats including evolving and intelligent attacks~\cite{hou2020low,brundage2018malicious, jang2014survey, camp2019measuring}. However, traditional defensive techniques cannot avoid the novel and evolving attacks which can leverage AI technology to plan and launch various attacks. AI-powered attacks can be categorized based on {\em AI-aided} and {\em AI-embedded} attacks. AI-aided attacks are those that leverage AI to launch the attacks effectively. In this type, the intelligent attackers use AI techniques. However, in AI-embedded attacks, the threats are weaponized by AI themselves such as Deep locker~\cite{stoecklin2018deeplocker} while in the AI-aided attacks, the attackers could launch various AI-based techniques to detect and recognize the target network, vulnerabilities, and valuable targets~\cite{kaloudi2020ai}. In fact, they utilize various AI techniques as a tool for various purposes. In~\cite{kaloudi2020ai}, the authors investigated the AI-powered cyber attacks and mapped them onto a proposed framework with new threats including the classification of several aspects of threats that use AI during the cyber-attack life cycle.

AI-powered threats are emerging attacks that use AI capabilities to launch various types of attacks to the system such as target attacks, botnet attacks~\cite{wang2020ai}, DeepLocker~\cite{stoecklin2018deeplocker}, and so on. Recently, the botnet has been studied in the literature as one of the most crucial threats today. {A botnet refers to a group of compromised computers that are remotely controlled by a botmaster via command and control (C\&C) channels~\cite{singh2019issues}}. Based on botnets, multiple types of cyber attacks can be launched against the clouds' infrastructures and resources from various contexts such as compromised IoT devices, sensors, or online social networks (OSNs)~\cite{faghani2019mobile}. Distributed Denial of Service (DDoS), Spam, Cryptocurrency Mining is some examples of botnet attacks. However, there are many defensive techniques to detect and defense against botnet attacks~\cite{xing2021survey}. The problem arises when the botnets leverage evolving technologies such as deep neural networks (DNN) to avoid detection and conceal the intention and targets. An example of intelligent attacks is the new generation of botnet attacks which are very difficult to be detected by the current decisive strategies~\cite{wang2020ai}.

Many studies have proposed novel techniques to enhance the strength of defensive techniques to defend against novel threats~\cite{kaloudi2020ai, wei2021ae, xu2021improving, zhu2020multi, zhu2021task}. However, they are still many AI-powered attacks that cannot be detected by strong IDS~\cite{hou2019intrusion}. To this end, it is crucial to evaluate the capabilities of both novel threats and possible strategies of defensive systems against the AI-powered attacks using strong decision-making models such as game theory models~\cite{shan2020game}.

Game Theory has proved to be very effective in cyber security areas and evaluation of defensive systems by taking some important decisions, devising plans, and optimizing the solutions and has been widely used to address the security of networks and systems~\cite{manshaei2013game,attiah2018game,rass2017physical}. {Game theory is a decision-responsive mathematical model which makes one side of a game change its strategy according to the decision made by the counterpart. To a certain extent, game theory can be defined as the principle for using mathematical models to study the conflict and cooperation between intelligent and rational decision-makers.} Security problems can be modeled between the players which are mostly a defender (i.e., IDS) and an attacker (i.e. external botnets). However, it is not always possible to determine if the traffic coming from the cloud users is normal or malicious~\cite{lin2012using}. Game theory enables the defender to model its interaction with users whose intentions are unknown to the IDS.

The main contributions of this paper are as follows:

\begin{itemize}
	    \item We provide a comprehensive case study that investigates the main characteristics and the core set of AI functionalities involved in the new generation of AI-based botnet attacks.
    \item We propose a sequential game theoretical model that formulates both attacker's and defender's best strategies and responses then solves the game based on Nash Equilibrium.
    \item In our proposed model, the utility function is computed based on the attacker's objective to launch the maximum number of DDoS attacks (i.e., to maximize the probability that the target system become unavailable) with the minimum attack cost while the defender's objective is to utilise the maximum number of defense strategies (i.e., to maximize the probability that the victim’s system recovers) with the minimum defense cost.
 \item We provide a numerical analysis on a simulated cloud environment with a various number of defense strategies on different cloud-band sizes against different attack success rate values. Our experimental results confirm that the success of botnet defense has a significant correlation with the attack rates.
\end{itemize}

The rest of the paper is organized as follows. Section 2 presents some background and related works. Section 3 provides the definitions, notations, and concepts used throughout the paper. In Section 4 Game Theory Model including the definition and notations as well as the game model is presented. The experimental Results and analysis are given in Section 5. Section 6 is the discussion and limitations. Finally, we conclude the paper in Section 7.

\begin{figure*}[t]
	\centering
	\includegraphics[width=1\linewidth]{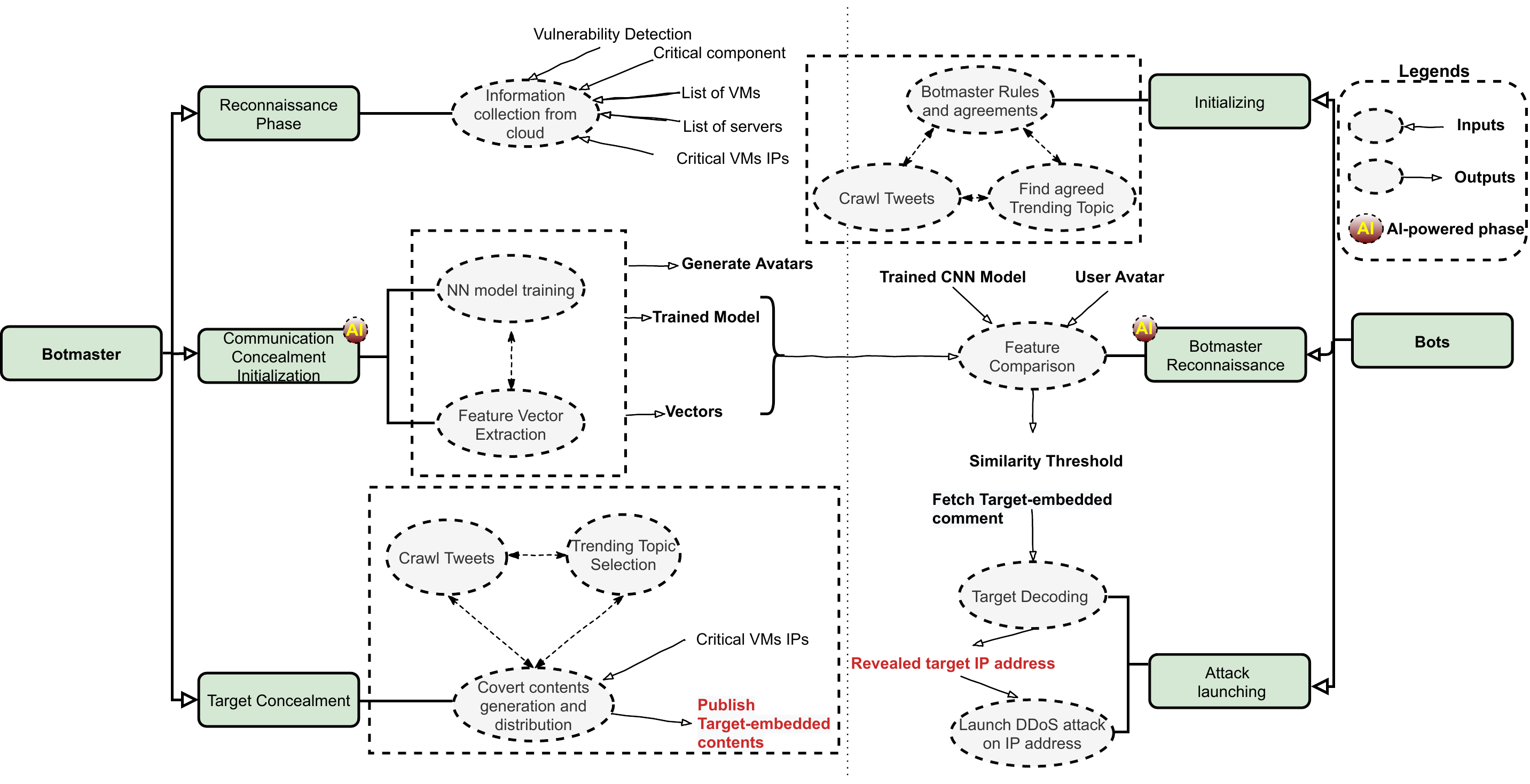}
	\caption{The main phases of botmaster and bots including detailed steps, requirements, and relations in the AI-powered botnet.}
	\label{fig:Botnet-phases}
\end{figure*}

\section{Related Work} \label{sec:rel} 
\subsection{Botnet Detection Techniques}
Bots can launch multiple types of cyberattacks especially Distributed Denial of Service (DDoS) attacks through a Command and Control (C\&C) channel which makes them able to communicate with a botmaster. In fact, the botmasters forward the constructive commands to the bots through the C\&C channel. Botnets utilize various methods to create C\&C channels between bots and botmaster. {Traditional C\&C channels are built using IRC, HTTP, P2P, and other protocols~\cite{pantic2015covert}.}

Various botnet detection techniques have been proposed in literature such as abnormal behavior~\cite{maimo2018self,nguyen2020psi}, communication signature-based techniques~\cite{singh2019issues}, DGA botnet detection~\cite{pei2020two}, and so forth~\cite{singh2019issues}.  In \cite{maimo2018self}, the authors proposed an anomaly detection system based on a botnet dataset through a deep learning system. In~\cite{guang2016using}, the authors utilized a deep learning method to detect bot activities on the cloud. Based on their method, the {basic features were extracted from the network stream and mapped onto grayscale images, Convolutional Neural Network (CNN) was adopted for feature learning, and SVM was used for classification detection}. In~\cite{guang2016using}, the authors proposed a botnet detection model based on the {channel information (such as power consumption). The information was collected on the side of the IoT, and CNN was used to perceive subtle differences in power consumption data.}

{Deep Neural Network (DNN) models have been widely used to detect botnet attacks~\cite{xing2021survey,vinayakumar2020visualized}. In~\cite{vinayakumar2020visualized}, the authors proposed a real-time botnet detection method using a two-level deep neural network. They utilized Siamese neural network to find out the similarity of DNS queries. Similarly, in~\cite{feng2017classification}, the authors used the Deep learning Image-Net model to classify DNS generated by DGA.  Another framework developed for botnet detection named Botnet Security System (BoNeSSy) was proposed in~\cite{gaonkar2020survey}. Their proposed method leverages DNN to identify valid and invalid botnet patterns. The advantage of the DNN-based technique is that it is good to identify the botnet traffic pattern. However, DNN-based methods usually suffer from run-time complexity during the learning phase. The run-time in DNN usually increases with the increase in the number of features. Moreover, most of the proposed methods cannot detect the new generations of botnets using AI-powered methods to conceal their intention and targets.} As the DNN models cannot be trained for detecting new botnets, it is difficult for them to identify their malicious activities.

{Previously, communication of bots and botmasters in an environment (such as IoT or OSNs) was established by hardcoding identification methods into the bots such as IDs, Links, or DGAs. The traditional methods of constructing a secure C\&C channel between bots and botmaster have some disadvantages. First, as the identification methods need to be hardcoded into bots, once the defender analyzed the bots, the C\&C communications can be monitored and detected. Consequently, the identity of the botmaster could be exposed to the defender. On the other hand, the commands passing through the C\&C communication channel are usually encrypted. These encrypted contents help the botnet detection methods to detect abnormal content (i.e. unnecessary encrypted message).
{With the evolution of Botnet detection, construction of C\&C channels pays more attention to concealment and began to utilize some public services such as social network~\cite{yin2018study}.}
However, in order to create an anomaly-resistant C\&C communication on a botnet environment (such as OSNs), In~\cite{pantic2015covert}, the authors proposed a method that embeds commands into tweets using metadata of tweets (length). However, the length of a tweet represents an ASCII code in decimal. As a tweet has a maximum length of 140, 7 bits can be conveyed through one tweet. While commands can be issued stealthily, the system has a low capacity. It needs to post N tweets to publish command in the length of N.}

\subsection{GT-based Botnet Evaluation}
Game theory has long been applied to study network security~\cite{chukwudi2017game} and game-theoretic models have been used for botnet detection and evaluation in various studies~\cite{asadi2021detecting,wang2016dynamic}.

The study of [10] demonstrates the exploitation pattern of an inherent weakness in local-host alert correlation-based methods and asserts that current local-host alert correlation implementations could allow pockets of cooperative bots to hide in an enterprise-size network. The study of [11] gives a graph-based representation of the infected computers. It provides that we can use graph-partitioning algorithms to separate out the different botnets when a network is infected with multiple botnets at the same time. The study of [12] gives a method for the detection of DDoS attacks using data mining. The study of [13] gives a brief detail of botnet DDoS attack behavior using queuing theory. In [14], a game theory-based model was presented as a defense mechanism against a classic bandwidth-consuming DoS/DDoS attack, which validates the model using NS-3 network simulation tools.

The study of [15] presents a general incentive-based method to model attacker intent, objectives, and strategies and a game-theoretic approach to inferring the ability to model and infer attacker intent, objectives, and strategies. Each attack host is considered as an attacker separately in[15]. In fact, if these hosts are infected and controlled by a botmaster, the threat of a DDoS attack, which is launched by the botnet, is much greater than one attack host.



However, it is important to consider how game theory can be applied to novel and emerging threats in cybersecurity~\cite{xing2021survey}.

\section{Threat Model on AI-powered Botnet}\label{sec:threats}


\subsection{Key technologies of AI-powered Botnet}
We investigate a new generation of botnets utilizing AI-embedded capabilities for launching various attacks toward a target system. The threat is equipped with AI capabilities which provide the bots and botmasters to be concealed and become undetectable through current botnet detection techniques~\cite{xing2021survey,guang2016using}. However, the new generation of botnets addressed this problem by using neural networks~\cite{wang2020ai} by covering the C\&C communication channel and generating command-embedded contents which seem normal (such as comments in social media, tweets, etc.). These two AI-powered features make the botnet capable to evade current detection techniques such as communication signature, abnormal behaviors, and so forth~\cite{xing2021survey}.

Later on in this section, we show how botmasters and bots can further communicate with each other and launch malicious DDoS attacks on a cloud system. They utilize NN to establish undetectable covert communication. Figure~\ref{fig:Botnet-phases} shows the main phases of both botmaster and bots from discovering the targets to launching DDoS attacks.

\subsection{Botmaster Main Phases}
\subsubsection{Reconnaissance Phase}
In this phase, the botmaster gain required information from the targeted cloud such as vulnerabilities, a list of critical VMs and servers, and the IP address of critical targets. Botmaster is able to use various reconnaissance tools and techniques such as Nessus, Open Vulnerability Assessment Scanner (OpenVAS) for vulnerability detection, and many network discovery and mapping tools such as IPsonar and Nmap~\cite{chauhan2018practical} to gain enough information about the target cloud. We later show that how the botmaster distributes the IP addresses of the targeted VMs in the OSN botnet.

\begin{figure}[t]
	\centering
	\includegraphics[width=0.98\linewidth]{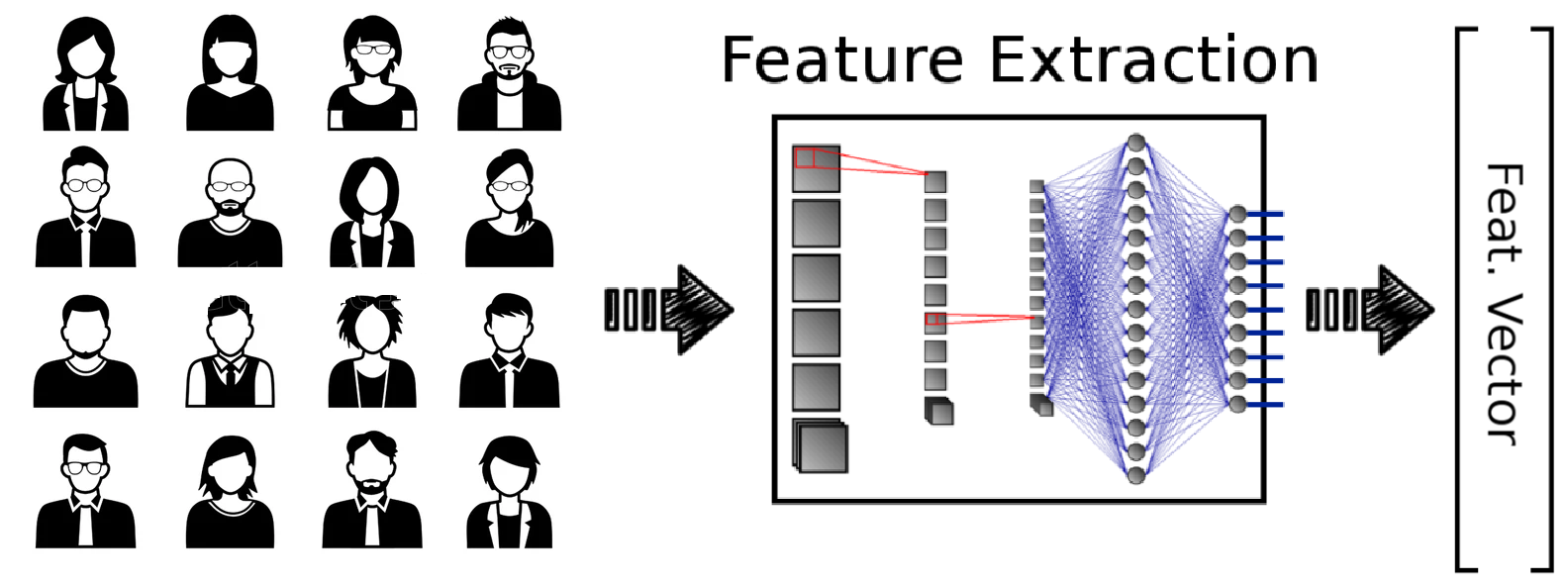}
	\caption{Vectors' feature extraction using CNN.}
	\label{fig:FE}
\end{figure}

\subsubsection{Communication Concealment Initialization Phase}
Botmaster needs to provide the bots with some criteria to discover the botmaster. As stated earlier, the current method of communication of bots and botmasters is mostly by hardcoding identification of botmasters such as IDs, Links, or DGAs methods into the bots which is detectable by current botnet detection methods~\cite{ghosh2021using}. To this end, botmaster needs to conceal its communication with bots. Thes, the neural network model can be used to protect accounts of botmaster and conceal the intent of bots. The bots are able to find the botmaster using the CNN model. {This phase includes two main steps: NN model training and feature vectors extraction (as in Figure~\ref{fig:FE}). First, the botmaster trains a NN model to extract feature vectors from some selected avatars. The high-dimensional features are built into bots so that bots can identify botmasters accurately when retrieving comments.}

\begin{figure}[t]
	\centering
	\includegraphics[width=0.98\linewidth]{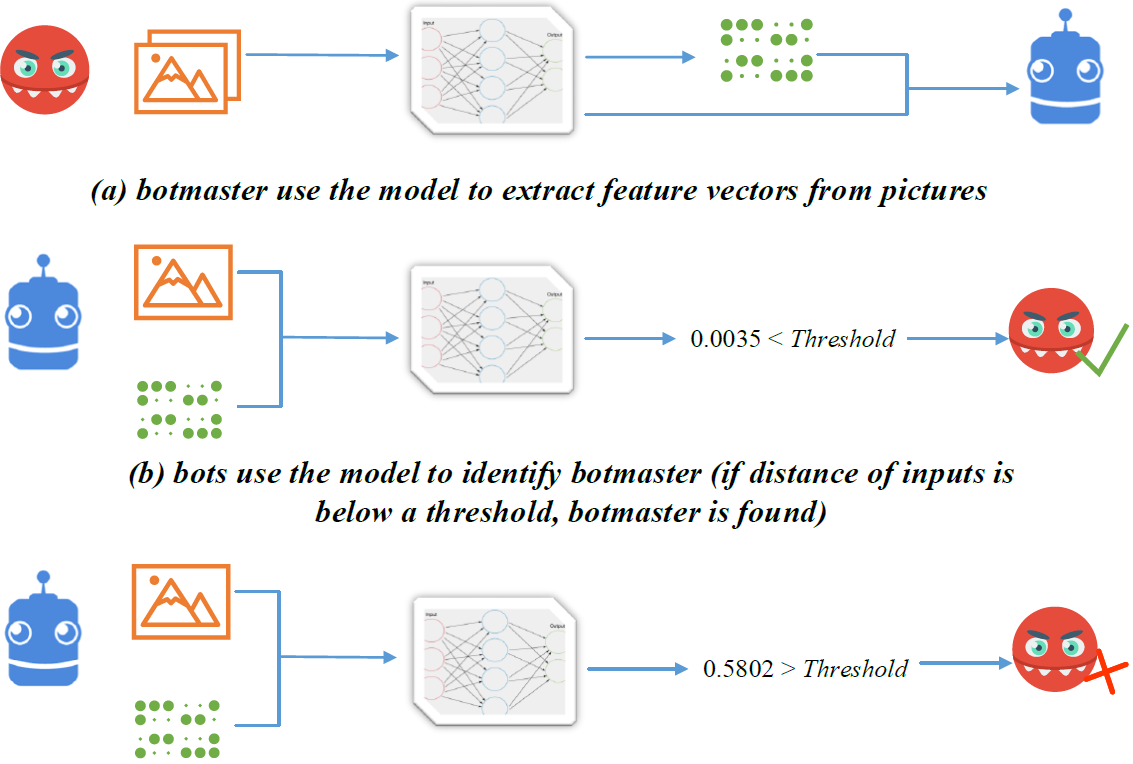}
	\caption{Finding botmaster using trained CNN model and selected features~\cite{wang2020ai}}
	\label{fig:botmaster-finding}
\end{figure}

\subsubsection{Target Concealment Phase}
In this step, the botmaster first needs to determine a trending tweet by crawling the tweets. Then, the botmaster designs a set of rules and agreements for Twitter trends and builds the bots with the rules, and distributed them along with the vectors and model. Moreover, the botmaster needs to generate the covert contents. The contents include the IP address of the targeted VM in the cloud system resulting from the reconnaissance phase of the botmaster. However, if the botmaster decides to encrypt the target IP address in the commands, the abnormal contents will raise anomalies on OSNs. Thus, the posted tweets need to have natural and semantic information. When the bots find the botmaster, the bots parse the comments by calculating the hashes of the tweets. Then, the target VM can be revealed.
In fact, the target IP addresses are embedded into tweets through hash collision. The Target-embedded tweets are posted by the botmaster. After bots found botmaster, IP addresses can be obtained by calculating hashes of the tweets. Thus, for launching attacks on various critical VMs in the cloud, the botmaster needs to generate as many as natural tweet contents that are about the trending topic. For this purpose, the botmaster needs to train a DNN model which requires a lot of data. Botmaster can use data augmentation techniques such as easy data augmentation (EDA) to enlarge the existing dataset and train the DNN model~\cite{WeiZ19}.
By this method, the botmaster can convey the IP address of target VMs to bots to launch DDoS attacks from the botnet to the denoted critical VMs.

\subsection{Bots Main Phases}
\subsubsection{Initializing}
{In this phase, while bots cannot find botmaster quickly among Twitter users without hardcoded rules, Twitter trends provide a meeting point for them. Twitter trends change with the volume of the tweets under different topics and are updated every 5 minutes, which is difficult to predict. Since ordinary users also discuss different topics, botmaster can hide among them. After the selection of trending topics and embedding of commands, the botmaster posts commands-embedded tweets to the trend. Bots also select a trending topic according to the agreed rule and crawl a lot of tweets along with tweeters' avatars to the trend. Then bots start identifying botmaster by its avatar.}

\subsubsection{Botmaster Reconnaissance}
{As botmaster's avatar is converted to a vector and distributed with bots. As both the vectors and the model are built into bots, bots can pick botmaster up quickly by comparing distances using the Convolutional Neural Network (CNN).} The trained model is used
to calculate the distances between avatars from Twitter uses and the built-in vectors to identify botmaster. A selected vector and a crawled avatar are fed into the model, and the model outputs the distance of the inputs. By comparing the distance with a threshold, bots can identify whether the avatar is from botmaster or not (as illustrated in Figure~\ref{fig:botmaster-finding}).

\subsubsection{Launching Attack}
After the botmaster is recognized by bots. The comments need to be parsed to obtain the IP address of the target VM in the cloud. Bots compute the hashes of tweets posted by the botmaster to find a successful collision and find the IP address. However, the advanced attackers can utilize AI techniques such as those as discussed above to launch various Distributed denial of service (DDoS) attacks to the target system (such as critical cloud components).

DDoS can be launched from various sources such as IoT and OSNs toward targets aiming to disrupt the crucial services in a network or the VMs running on a cloud. This can lead to service and system disruption and finally causes system breakdown. After finding the targets, botnet launches targeted VMs remotely (such as in ~\cite{kolias2017ddos}). The well-known botnets such as Mirai~\cite{kolias2017ddos} can be detected by botnet detection techniques. However, a new generation of botnets explained in this section using AI-powered technologies equipped with target and intention concealment cannot be detected by current detection methods.

Thus, it is important to utilize the game theory to model the attacker and defender's capabilities and evaluate the best strategies for both parties against each other. This may help the defense to find its ultimate defensive strategies against such attacks. Moreover, the game theory models can be leveraged to evaluate the effects of these types of attacks against the target system (e.g. cloud system).

\begin{figure*}[t]
	\centering
	\includegraphics[width=0.90\linewidth]{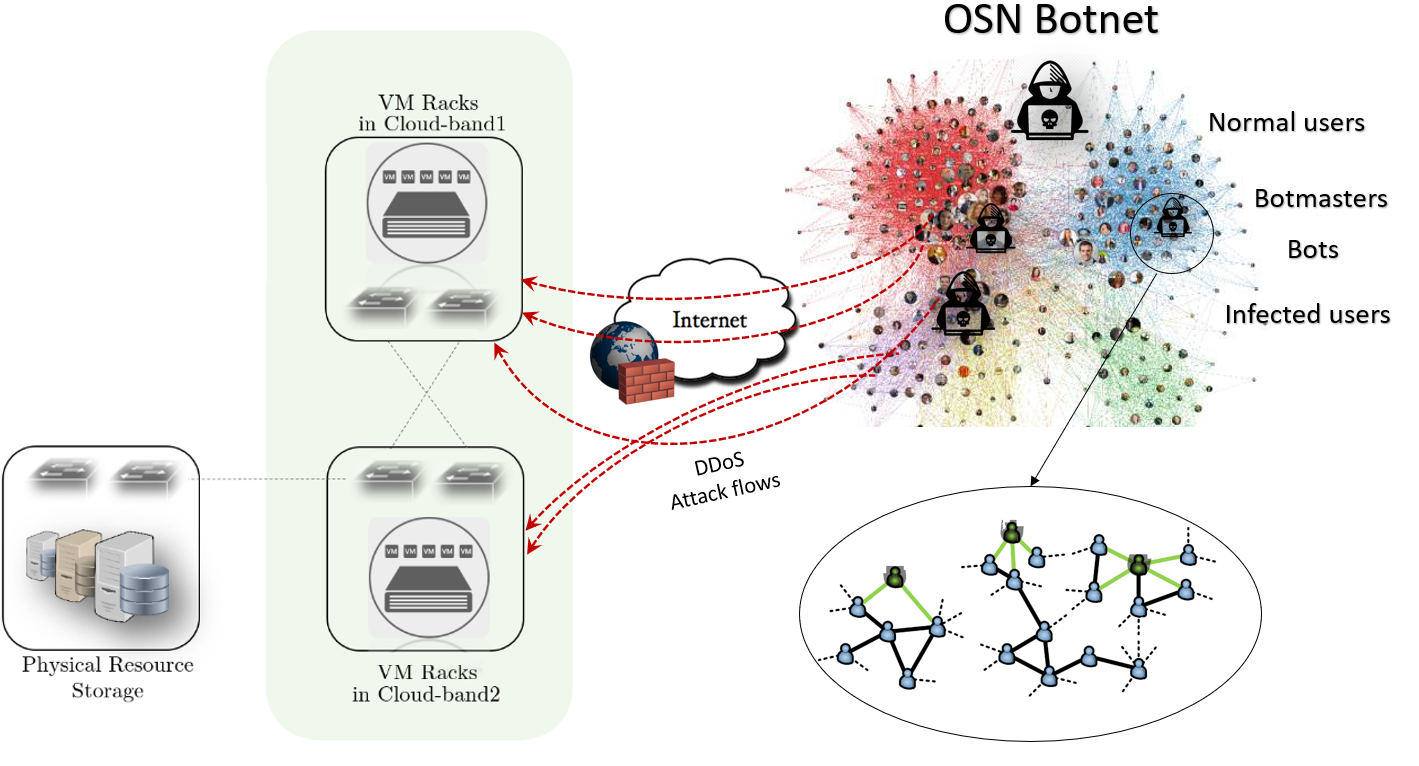}
	\caption{System Model: DDoS attacks flow from Online Social Networks (OSN) Botnet toward the Cloud VMs.}
	\label{fig:MK}
\end{figure*}

\section{DDoS Botnet Attack Evaluation Through Sequential Game Theory Model}\label{sec:GT}

\subsection{System Model}
We define our model based on a cloud system including multiple cloud-band nodes that are connected to each other hierarchically. We first define a cloud system as follows.
\theoremstyle{definition}

\begin{definition}\label{CH6:CS}
	A cloud system can be defined as $\mathcal{C}=\{\mathcal{N},E\}$, where $\mathcal{N}$ is a set of cloud components or nodes including Virtual Machines (VMs), routers, switches, and so forth. Then, $E$ is a set of connections between the cloud's components. Then, the cloud system includes $n$ number of sub-clouds or cloud-bands $CB=\{\hat{c}_1, \hat{c}_2, \dots, \hat{c}_n\}$. We then define a cloud-band as $\hat{c}_k=\{VM_k,E_k\}$ where $VM_k$ is a subset of VMs in cloud-band $\hat{c}_k$ and $E_k\subseteq \{VM_k \times VM_k\}$ denoted the connectivities of those VMs in cloud-band $\hat{c}_k$.
\end{definition}



\subsection{Notations and Assumptions}
The notations and definitions used for modeling attack and defense are presented as follows:
\begin{itemize}
    \item $\hat{c}_1$: cloud-band 1, main cloud entry, at level 1.
    \item $\hat{c}_2$: cloud-band 2, cloud node, at level 2.
    \item $\hat{c}_3$: cloud-band 3 or resource node at level 3.
    \item $n$: the total number of cloud-bands that are connected to each other hierarchically. In here $n=3$, as the cloud system has only three cloud-band or cloud levels.
    \item $|\hat{c}_k|$: the total number of VMs existing in a specific cloud-band $\hat{c}_k$.
    \item $min_k$: minimum number of active, defended, or un-attacked VMs for a specific cloud-band $\hat{c}_k$ so that the cloud-band can maintain its functionality (providing reliable service to the users).
    \item $\alpha_k$: the number of VMs in cloud-band $\hat{c}_k$ which are under botnet attacks.
    \item $e_k$: a subset of VMs in cloud-band $\hat{c}_k$ which are fully exploited by flowing DDoS Botnet attack. Note that those VMs lose their functionality due to successful attack. Note that $|e_k|\leq |\hat{c}_k|$.
    \item $\delta_k$: the number of VMs recovered or defended VMs in cloud-band $\hat{c}_k$. Note that those VMs gain their functionality due to successful defense.
	\item $p_{k}$: the probability of attack success for each VM in cloud-band $k$ (VM is exploited or under DDoS attack). Note that in the presence of successful attack, the VM fails to serve its task.
	\item $L_{d,k}$: the defender loss for cloud-band $\hat{c}_k$ resulting from the breakdown of that specific cloud-band. However, $L_{d,k}$ shows the importance of cloud-band $\hat{c}_k$. Thus, higher value of $L$ indicated higher importance of cloud-bank.
	\item $G_{a,k}$: the attacker gain for cloud-band $\hat{c}_k$ resulting from the breakdown of that specific cloud-band.
	\item $f_{k}$: the probability that attacker launch internal attack from exploited VM in cloud-band $\hat{c}_k$ to adjacent cloud-band $\hat{c}_{k+1}$ or $\hat{c}_{k-1}$.
	\item $C_{d,k}$: the cost of a successful defense or recovery of exploited VM in cloud-band $\hat{c}_k$. Note that the cost unit is based on each VM.
	\item $C_{a,k}$: the cost of a successful attack in cloud-band $\hat{c}_k$. Note that the cost unit is based on each VM.
	\item $P_k(e_k,\delta_k)$: the probability that cloud-band $\hat{c}_k$ is functional (the cloud-band provide reliable services to the clients) under the number of attacks $\alpha_k$ and defense $\delta_k$ for cloud-band $\hat{c}_k$.
\end{itemize}


\section{Game Model Formulation}\label{sec:GT}
{We modeled a sequential game where the defender movement is before the attacker's actions. In practice, a sequential game where the defender moves first and might be more realistic than a simultaneous game. The defender could intentionally reveal the defense strategy in order to influence the attacker's strategy or the attacker could use techniques such as spying to learn about the defender's strategy~\cite{sanjab2016bounded}.}

We define the sequential game as $G=<(\mathcal{P}),(\mathcal{S}),(\mathcal{U})>$, where $\mathcal{P}=\{\mathcal{A}, \mathcal{D}\}$ denotes two players as attacker and defender. The attacker is the malicious botnet attacker which uses various bots to execute it's commands over the cloud, and the other player is the cloud defender. Both attackers and defenders are two players of a game and can be represented by $\mathcal{A}$, and $\mathcal{D}$, respectively. Each player ($p\in \mathcal{P}$) can choose a (pure) \textit{Strategy} $s_p$ from the \textit{strategy space} $S_p=\{s_p^1,....,s_p^j\}$. The strategy for the defender is to assign a number of defense for each cloud-band $\hat{c}_k$. Note that assigning a defense incurs cost for the defender. Thus, the defender should find reasonable number of defenses for each cloud-band to gain most security with lower defense cost. In fact, for the cloud system which includes three cloud-bands, the defense strategy is to devote a number of defense to each cloud-band to avoid the sub-cloud's breakdown with the lowest cost. For instance, $s_\mathcal{D}^i=(\delta_{\hat{c}_1},\delta_{\hat{c}_2},\delta_{\hat{c}_3})$ indicates that the defender deployed $\delta_{\hat{c}_1}$, $\delta_{\hat{c}_2}$, and $\delta_{\hat{c}_3}$ number of defenses to cloud-bands $\hat{c}_1$, $\hat{c}_2$, and $\hat{c}_3$, respectively. However, the attacker objective is to launch as many as DDoS attacks to the VMs in the cloud system to maximize the probability that system become unavailable with minim attack effort and cost. Then, the attacker's strategy can be defined by assigning the number of attacks to each cloud-band. For instance, $s_\mathcal{A}^i=(\alpha_{\hat{c}_1},\alpha_{\hat{c}_2},\alpha_{\hat{c}_3})$ indicates that the attacker launched $\alpha_{\hat{c}_1}$, $\alpha_{\hat{c}_2}$, and $\alpha_{\hat{c}_3}$ number of attacks to cloud-bands $\hat{c}_1$, $\hat{c}_2$, and $\hat{c}_3$, respectively. Thus, $s_\mathcal{A}^1=(\alpha_{\hat{c}_1}=5,\alpha_{\hat{c}_2}=5,\alpha_{\hat{c}_3}=2)$ indicates that the first attacker's strategy is to launch 5 DDoS attacks to cloud-band $\hat{c}_1$, together with 5 and 2 attacks on cloud-bands $\hat{c}_2$, and $\hat{c}_3$, respectively.

The \textit{outcome} of the game is a final strategy chosen by each player as $s=(s_{\mathcal{A}},s_{\mathcal{D}})$. The \textit{payoff function} for a player for a given strategy can be defined as $u_p(s_{\mathcal{A}};s_{\mathcal{D}})$, and $S$ is a set of all possible strategies that can be chosen by both attacker and defender such that $S\in S_{\mathcal{A}} \times S_{\mathcal{D}}$.

\subsection{Utility Function}\label{sec:UF}
{In this game, the defender makes effort to find out a trade-off between the defense costs and security benefits. However, the attacker intends to maximize the breakdown probability of each cloud band with a reasonable attack cost.}

{We first study the probability the cloud-band $\hat{c}_k$ works properly under attacks. The attacker launch botnet DDoS attacks toward the VMs of each cloud-band, especially cloud-band $\hat{c}_1$, while the defender tries to defend a number of VMs in the cloud bands. In the presence of successful defense, the impacted VM can be recovered. The probability that the cloud-bands preserve their functionality can be modeled based upon a binomial random variable, which is defined under the number of failed VMs due to successful attacks (under DDoS attacks). In this case, the probability describes a binomial distribution, where the probability of success is the combined probability of a successful attack on a VM of cloud-bank $\hat{c}_k$. Thus, the probability of functionality of the cloud-band $\hat{c}_k$ is 0 if $0\leq \delta_k<min_k$ as there are fewer defense efforts against the minimum number of operational VMs, and $\hat{c}_k$ is 1 if $\delta_k\ge e_k+min_k$ which means the cloud-band will operate normally since the defense efforts is more than the sum of both attack efforts and required defense. Likewise, the probability of functionality of the cloud-band can be computed as Equation~\eqref{eq:PO}, if $min_k \leq \delta_k < e_k + min_k$.}

\begin{equation}\label{eq:PO}
\begin{split}
P_k(\alpha_k,\delta_k)=
  \sum_{|e_k|=0}^{\delta_k-min_k+1} {{\alpha_k}\choose{|e_k|}} p_k^{|e_k|}(1-p_k)^{\alpha_k-|e_k|}
\end{split}
\end{equation}


{Finally, the optimization problem can be formulated based on the objective functions of both attacker and the defender. Note that, the unit costs of defending and attacking would prevent unlimited defense and attack efforts. Consequently, we expect both the defender and attacker to try to find out a balance between defending and attack so that the attacker increases the breakdown probability of each cloud band with the lowest attack cost while the defender increases the functionality probability of the cloud bands with the lowest defense cost.} Based on the given objectives, the utility function of the defender aiming to defend the whole cloud system can be formulated as Equation~\eqref{eq:def-ob}.

\begin{equation}\label{eq:def-ob}
\begin{split}
u_\mathcal{D}(s_\mathcal{D};s_\mathcal{A})= u_\mathcal{D}(\alpha_{\hat{c}_1},\alpha_{\hat{c}_2},\dots,\alpha_{\hat{c}_n};\delta_{\hat{c}_1},\delta_{\hat{c}_2},\dots,\delta_{\hat{c}_n})=  \\
\max\limits_{\delta_{\hat{c}_1},\delta_{\hat{c}_2},\delta_{\hat{c}_3}} \underbrace{\sum_{k=1}^{n} P_{\hat{c}_k}L_{d,\hat{c}_k}}_\text{Defender gain for cloud's functionality}-
\underbrace{\sum_{k=1}^{n} C_{d,\hat{c}_k}\delta_{\hat{c}_k}}_\text{Total cost of defense}-\\
\underbrace{(1-P_{\hat{c}_1})(f_{\hat{c}_1,\hat{c}_2}L_{d,\hat{c}_2}+f_{\hat{c}_1,\hat{c}_2}f_{\hat{c}_2,\hat{c}_3}L_{d,\hat{c}_3})}_\text{Loss due to internal attack extension}
\end{split}
\end{equation}

Thus, for the cloud system represented in Figure~\ref{fig:MK}, the defender's utility function can be determined as Equation~\eqref{eq:def-ob-cloud}.

\begin{equation}\label{eq:def-ob-cloud}
\begin{split}
u_\mathcal{D}(\alpha_{\hat{c}_1},\alpha_{\hat{c}_2},\alpha_{\hat{c}_3};\delta_{\hat{c}_1},\delta_{\hat{c}_2},\delta_{\hat{c}_3})=  \\
\max\limits_{\delta_{\hat{c}_1},\delta_{\hat{c}_2},\delta_{\hat{c}_3}} P_{\hat{c}_1}L_{d,\hat{c}_1}+P_{\hat{c}_2}L_{d,\hat{c}_2}+P_{\hat{c}_3}L_{d,\hat{c}_3}-\\
C_{d,\hat{c}_1}\delta_{\hat{c}_1}-C_{d,\hat{c}_2}\delta_{\hat{c}_2}-C_{d,\hat{c}_2}\delta_{\hat{c}_2}-\\
(1-P_{\hat{c}_1})(f_{\hat{c}_1,\hat{c}_2}L_{d,\hat{c}_2}+f_{\hat{c}_1,\hat{c}_2}f_{\hat{c}_2,\hat{c}_3}L_{d,\hat{c}_3})
\end{split}
\end{equation}

Similarly, the utility function of the attacker aiming to attack the whole cloud system can be formulated as Equation~\ref{eq:att-ob}.

\begin{equation}\label{eq:att-ob}
\begin{split}
u_\mathcal{A}(s_\mathcal{D};s_\mathcal{A})=u_\mathcal{A}(\alpha_{\hat{c}_1},\alpha_{\hat{c}_2},\dots,\alpha_{\hat{c}_n};\delta_{\hat{c}_1},\delta_{\hat{c}_2},\dots,\delta_{\hat{c}_n})=  \\
\max\limits_{\alpha_{\hat{c}_1},\alpha_{\hat{c}_2},\alpha_{\hat{c}_3}} \underbrace{\sum_{k=1}^{n} (1-P_{\hat{c}_k})G_{a,\hat{c}_k}}_\text{Attacker's gain for cloud's breakdown}-
\underbrace{\sum_{k=1}^{n} C_{a,\hat{c}_k}\alpha_{\hat{c}_k}}_\text{Total cost of attacks}+\\
\underbrace{(1-P_{\hat{c}_1})(f_{\hat{c}_1,\hat{c}_2}L_{a,\hat{c}_2}+f_{\hat{c}_1,\hat{c}_2}f_{\hat{c}_2,\hat{c}_3}L_{a,\hat{c}_3})}_\text{Gain due to internal attack extension}
\end{split}
\end{equation}

Consequently, for the cloud system represented in Figure~\ref{fig:MK}, the attacker's utility function can be determined as Equation~\eqref{eq:att-ob-cloud}.

\begin{equation}\label{eq:att-ob-cloud}
\begin{split}
u_\mathcal{A}(\alpha_{\hat{c}_1},\alpha_{\hat{c}_2},\alpha_{\hat{c}_3};\delta_{\hat{c}_1},\delta_{\hat{c}_2},\delta_{\hat{c}_3})= \max\limits_{\alpha_{\hat{c}_1},\alpha_{\hat{c}_2},\alpha_{\hat{c}_3}}\\ {(1-P_{\hat{c}_1})G_{d,\hat{c}_1}+(1-P_{\hat{c}_2})G_{d,\hat{c}_2}+(1-P_{\hat{c}_3})G_{d,\hat{c}_3}}-\\
{C_{a,\hat{c}_1}\alpha_{\hat{c}_1}-C_{a,\hat{c}_2}\alpha_{\hat{c}_2}-C_{a,\hat{c}_2}\alpha_{\hat{c}_2}}+\\
{(1-P_{\hat{c}_1})(f_{\hat{c}_1,\hat{c}_2}G_{a,\hat{c}_2}+f_{\hat{c}_1,\hat{c}_2}f_{\hat{c}_2,\hat{c}_3}G_{a,\hat{c}_3})}
\end{split}
\end{equation}

\subsection{Game quantification and solving}\label{sec:nash}
\subsubsection{Nash Equilibrium}
Nash Equilibrium is used in the cyber security game theory to find the best response (defense) based on the \textit{mixed strategy}. Both defender and an attacker seek optimal strategies, and no one has an incentive to deviate unilaterally from their equilibrium strategies despite their conflict for security objectives. We denote a mixed strategy for a player $p$ as $\sigma_p = (p^1,...,p^n)$. Note that $\sum{_j}P^j=1$. The a set of all possible mixed strategies can be represented as $\Delta S_p$ for the player $p$ (the set of all probability distributions over the pure strategies of player $p$). Then, $\sigma=(\sigma_{p_{P_A}},\sigma_{p_{P_D}})$. Lets denote the $\sigma_{-p}$ as the mixed strategy of the opponent player (can be either attacker or defender). The the outcome of the game based on the mixed strategy can be written as $\sigma=(\sigma_p,\sigma_{-p})$. Then, the expected payoff to player $p$ as a function of the mixed strategy profile played by the players in the game can be represented as $u_p(\sigma_p,\sigma_{-p})$. A strategy profile $\sigma$ is a Nash Equilibrium if, for each $p$, her choice $\sigma_p$ is a best response to the other players' choices $\sigma_{-p}$, or based on math definition $u_p(\sigma_p,\sigma_{-p}) \ge u_p(s'_p,\sigma_{-p})$ for all $s'_p \in S_p$.

\subsubsection{Sub-game perfect Nash equilibrium}
In this sequential game, we assume that the defender moves first. The defender strategy may be revealed by the attacker. This helps the defender to evade the attacker and influence the attacker's strategy. We define the defender's best strategy as:

$$
s'_\mathcal{D}=\text{arg}\max_{\delta'_{\hat{c}_1},\delta'_{\hat{c}_2},\delta'_{\hat{c}_3}}u_\mathcal{D}(\alpha_{\hat{c}_1},\alpha_{\hat{c}_2},\dots,\alpha_{\hat{c}_n};\delta_{\hat{c}_1},\delta_{\hat{c}_2},\delta_{\hat{c}_3}),
$$
where $s'_\mathcal{D}=(\delta'_{\hat{c}_1},\delta'_{\hat{c}_2},\delta'_{\hat{c}_3})$ is the best response to the strategy used by the attacker $s'_\mathcal{A}=(\alpha_{\hat{c}_1},\alpha_{\hat{c}_2},\alpha_{\hat{c}_3})$.

However, we define the attacker's best strategy as:
$$
s'_\mathcal{A}=\text{arg}\max_{\alpha_{\hat{c}_1},\alpha_{\hat{c}_2},\alpha_{\hat{c}_3}}u_\mathcal{A}(\alpha_{\hat{c}_1},\alpha_{\hat{c}_2},\dots,\alpha_{\hat{c}_n};\delta_{\hat{c}_1},\delta_{\hat{c}_2},\delta_{\hat{c}_3}),
$$
where $s'_\mathcal{A}=(\alpha_{\hat{c}_1},\alpha_{\hat{c}_2},\alpha_{\hat{c}_3})$ is the best response to the strategy used by the defender $s'_\mathcal{D}=(\delta'_{\hat{c}_1},\delta'_{\hat{c}_2},\delta'_{\hat{c}_3})$.

In this section, we formulate the game theory model to evaluate the various strategies and interactions between both the defender and attacker in terms of assigning the number of attacks and defense against the three cloud layers (named as cloud-bands). We formulate both attacker's best DDoS attack strategies toward all cloud bands and the defender's best responses based on the function where the number of VMs attacked and the number of defense assigned to avoid the cloud breakdown.

\subsection{Numerical Results Analysis}

To conduct the experiments, we simulated a large cloud-band model as represented in Figure~\ref{fig:MK} as used in~\cite{alavizadeh2021evaluating, alavizadeh2019automated, alavizadeh2018evaluation, alavizadeh2017effective}. We assumed that the botnet DDoS attacks are launching from the online social network based on various attack rates as defined in Section~\ref{sec:threats}. The cloud model includes two cloud-band nodes that can hold up to 100 VMs each and a resource node. Each cloud band includes up to 50 critical VMs which are connected to the Internet (i.e., front-end servers). The exploitation of a portion of the critical VMs may cause cloud breakdown. We also assumed that the successful attack rates noted as the attack success probability (ASP) are the same for all critical nodes. The ASP denotes the severity of attacks launching through the OSN botnet on the cloud.

We study the attacker's and defender's responses assuming that the attacker can launch attacks to $\alpha_{k}$ number of critical VMs in each cloud-band $\hat{c}_k \in CB$. The defender can deploy up to $|\hat{c}_k|$ defense on each cloud band. We conducted a sensitivity analysis
to evaluate the effects of different key parameters on the proposed model. Table~\ref{param} represented the baseline parameters we utilized for conducting results.

We conduct the numerical results based on various number of defense on cloud-band $\hat{c}_1$ ($20\le\delta_{\hat{c}_1}\le 42 )$ based on various attack success rate $p_k$ values such as $0.1, 0.15, 0.20, 0.25$. Moreover, we evaluate the cloud functionality probability ($P_k$) based on various values of $p_k$ and minimum number of maintenance $min_k$ against attack number $\alpha_{\hat{c}_1}=30$. We further evaluated cloud functionality probability ($P_k$) based on various values of $p_k$ and minimum number of maintenance $min_k$ against attack number $\alpha_{\hat{c}_1}=50$.

\emph{Scenario (\romannumeral 1).} Figure~\ref{c01} depicts the results of cloud functionality probability values for a cloud that needs at least 20 running VMs. The results are conducted based on 30 attacks and a various number of devoted defense under four attack success rates. The results show that the appropriate number of defense for $P_k$ values of 0.1, 0.15, 0.2, 0.25 are 24, 26, 28, 30 respectively to keep the cloud system functional with the probability of higher $90\%$. It is also noticeable that devoting more than 30 defense has no effects on cloud functionality and only increases the defense cost.

\begin{table}[t]
\caption{Parameters and values used for conducting experiments}
\label{param}
\resizebox{0.49\textwidth}{!}{
\begin{tabular}{|l|l|l|}
\hline
Parameter & Details & Values \\ \hline
$|CB|$   & The number cloud-bands (levels)   & 1                          \\ \hline
$|\hat{c}_1|$   & The number of VMs in cloud-band 1    & 100                          \\ \hline
$min_{\hat{c}_1}$   & Minimal maintenance/defense for cloud-band 1               & 20--30                        \\ \hline
$\delta_{\hat{c}_1}$   & The number of defenses for cloud-band 1       & 20--42                         \\ \hline
$\alpha_{\hat{c}_1}$   & The number of attacks on critical nodes in $\hat{c}_1$              & 30, 50                        \\ \hline
$p_{\hat{c}_1}$   & Attack success probability for each VM in $\hat{c}_1$             & 10--25$\%$                       \\ \hline

$L_{d,1}$   & Defender loss for $\hat{c}_1$ breakdown            & 20                       \\ \hline
$G_{a,1}$   & Attacker gain for $\hat{c}_1$ breakdown            & 15                       \\ \hline
$C_{d,1}$   & Cost of defense for a critical VM in $\hat{c}_1$   & $0.5$    \\ \hline
$C_{a,1}$   & Cost of attack for a critical VM in $\hat{c}_1$   & $0.4$ \\ \hline
\end{tabular}}
\end{table}

	\begin{figure*}[t]
	\centering
	\begin{subfigure}{0.32\textwidth}
		\includegraphics[width=0.99\textwidth]{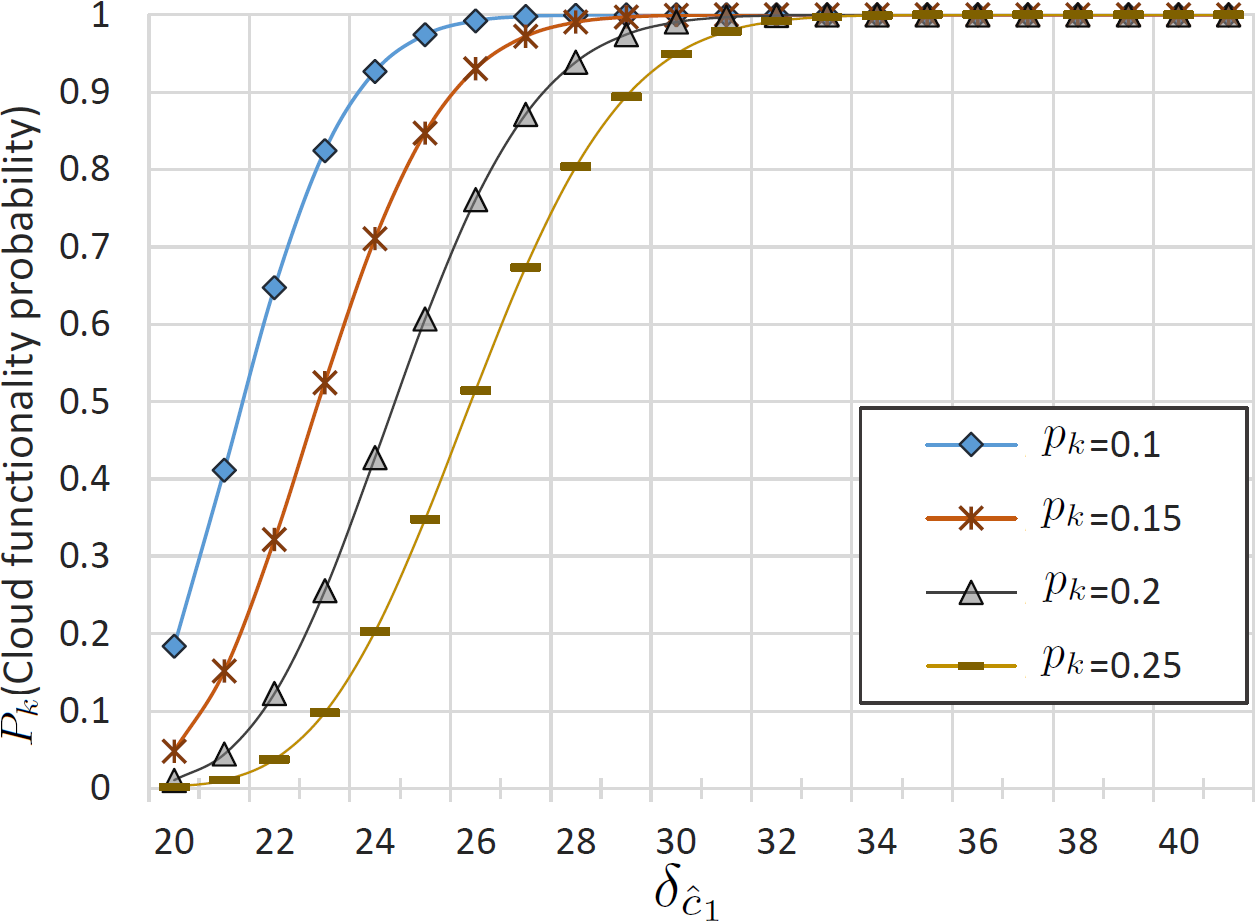}
		\caption{$P_{\hat{c}_1}(\alpha_{\hat{c}_1}=30,\delta_{\hat{c}_1}),~min_{\hat{c}_1}=20$}
		\label{c01}
	\end{subfigure}
	\begin{subfigure}{0.32\textwidth}
		\includegraphics[width=0.99\textwidth]{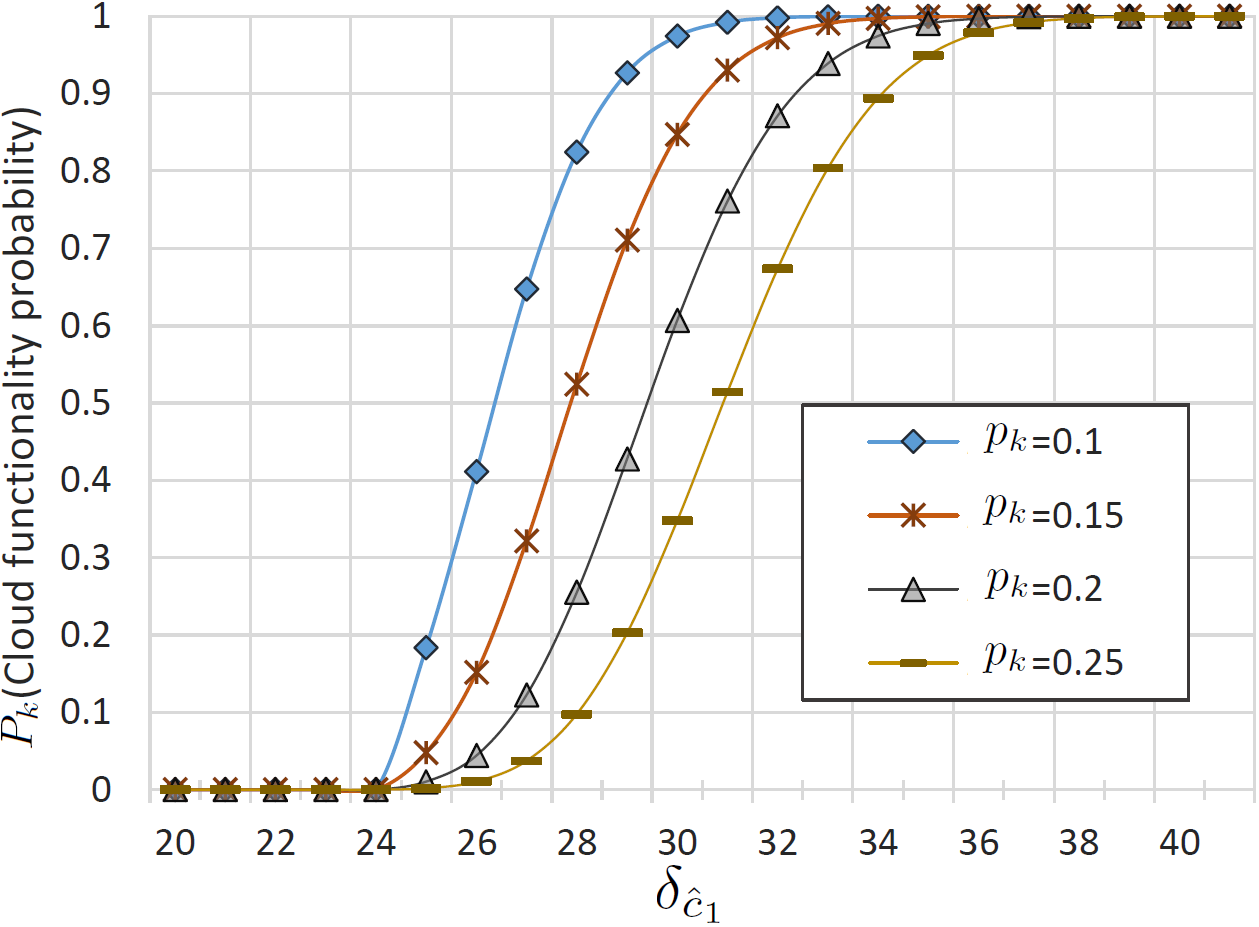}
		\caption{$P_{\hat{c}_1}(\alpha_{\hat{c}_1}=30,\delta_{\hat{c}_1}),~min_{\hat{c}_1}=25$}
		\label{c02}
	\end{subfigure}
	\begin{subfigure}{0.32\textwidth}
		\includegraphics[width=0.99\textwidth]{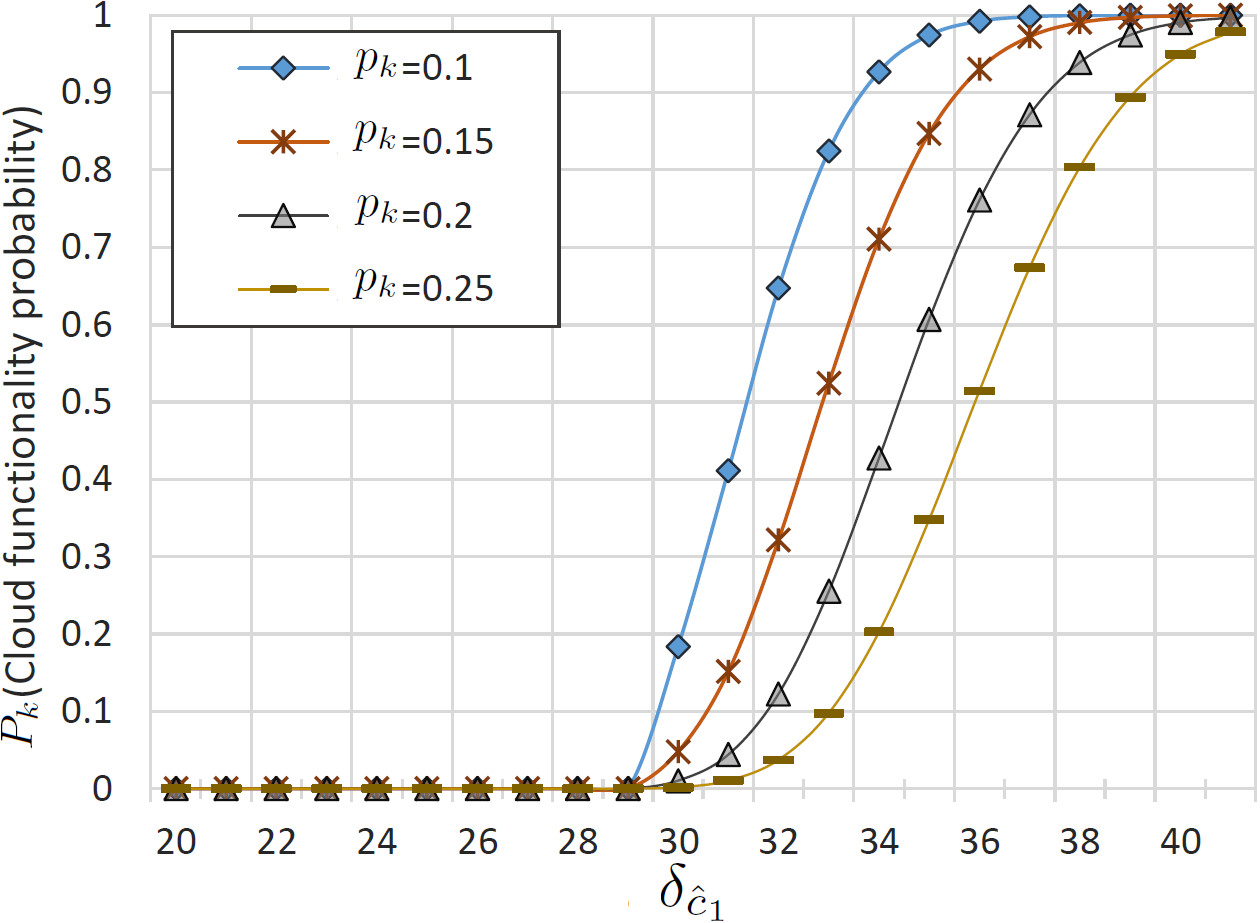}
		\caption{$P_{\hat{c}_1}(\alpha_{\hat{c}_1}=30,\delta_{\hat{c}_1}),~min_{\hat{c}_1}=30$}
		\label{c03}
	\end{subfigure}
	\\
	\vspace{3mm}
	\begin{subfigure}{0.32\textwidth}
		\includegraphics[width=0.99\textwidth]{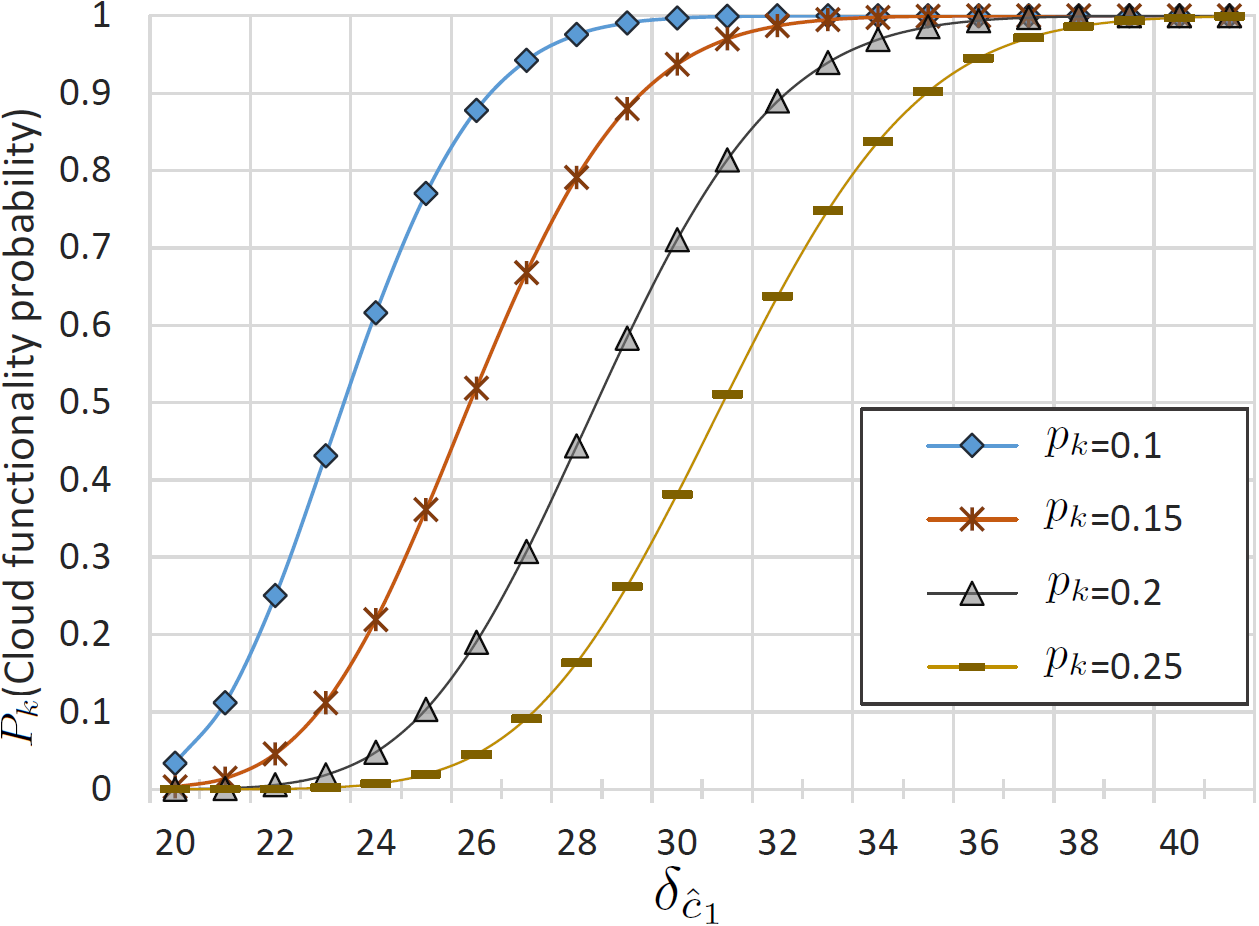}
		\caption{$P_{\hat{c}_1}(\alpha_{\hat{c}_1}=50,\delta_{\hat{c}_1}),~min_{\hat{c}_1}=20$}
		\label{a01}
	\end{subfigure}
	\begin{subfigure}{0.32\textwidth}
		\includegraphics[width=0.99\textwidth]{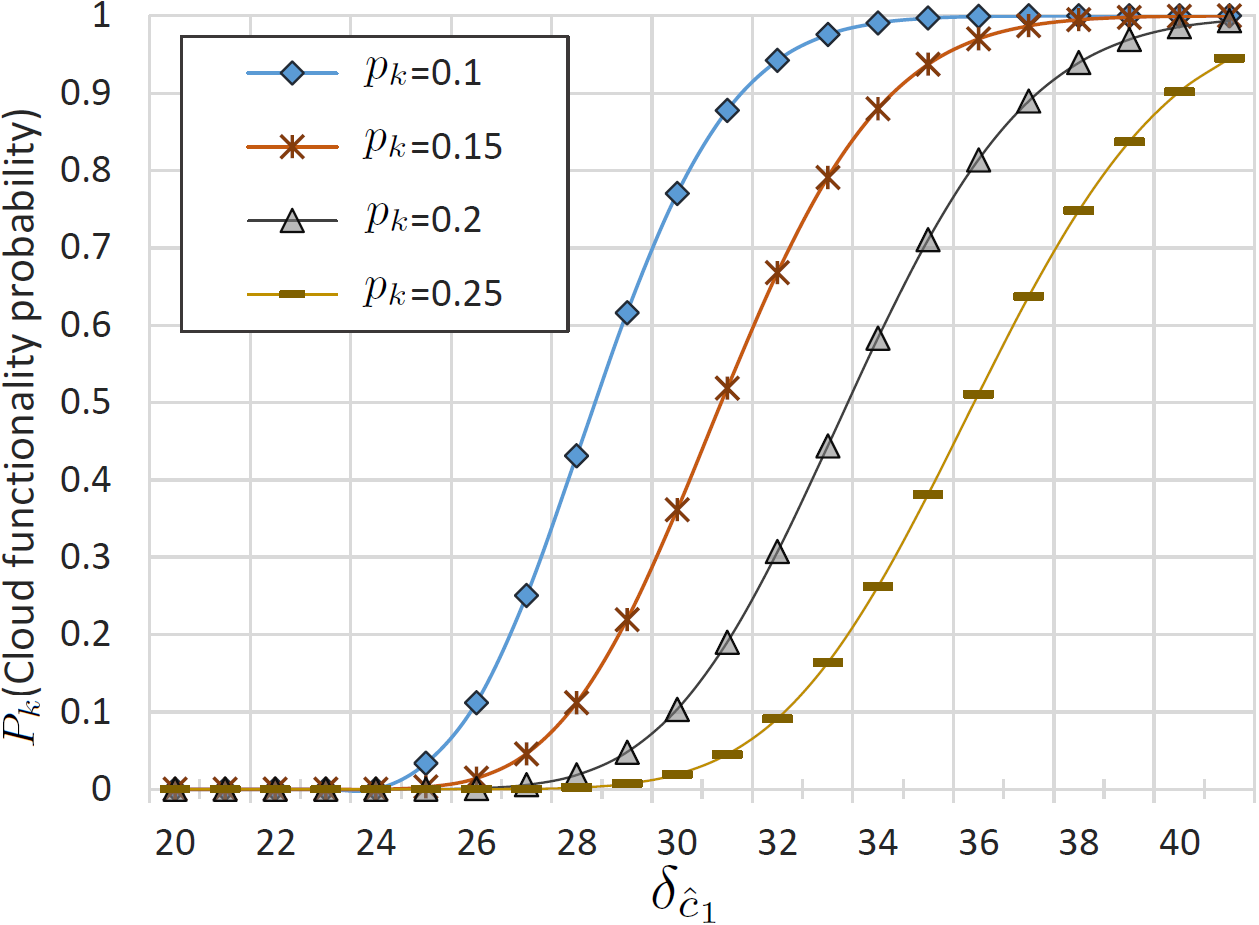}
		\caption{$P_{\hat{c}_1}(\alpha_{\hat{c}_1}=50,\delta_{\hat{c}_1}),~min_{\hat{c}_1}=25$}
		\label{a02}
	\end{subfigure}
	\begin{subfigure}{0.32\textwidth}
		\includegraphics[width=0.99\textwidth]{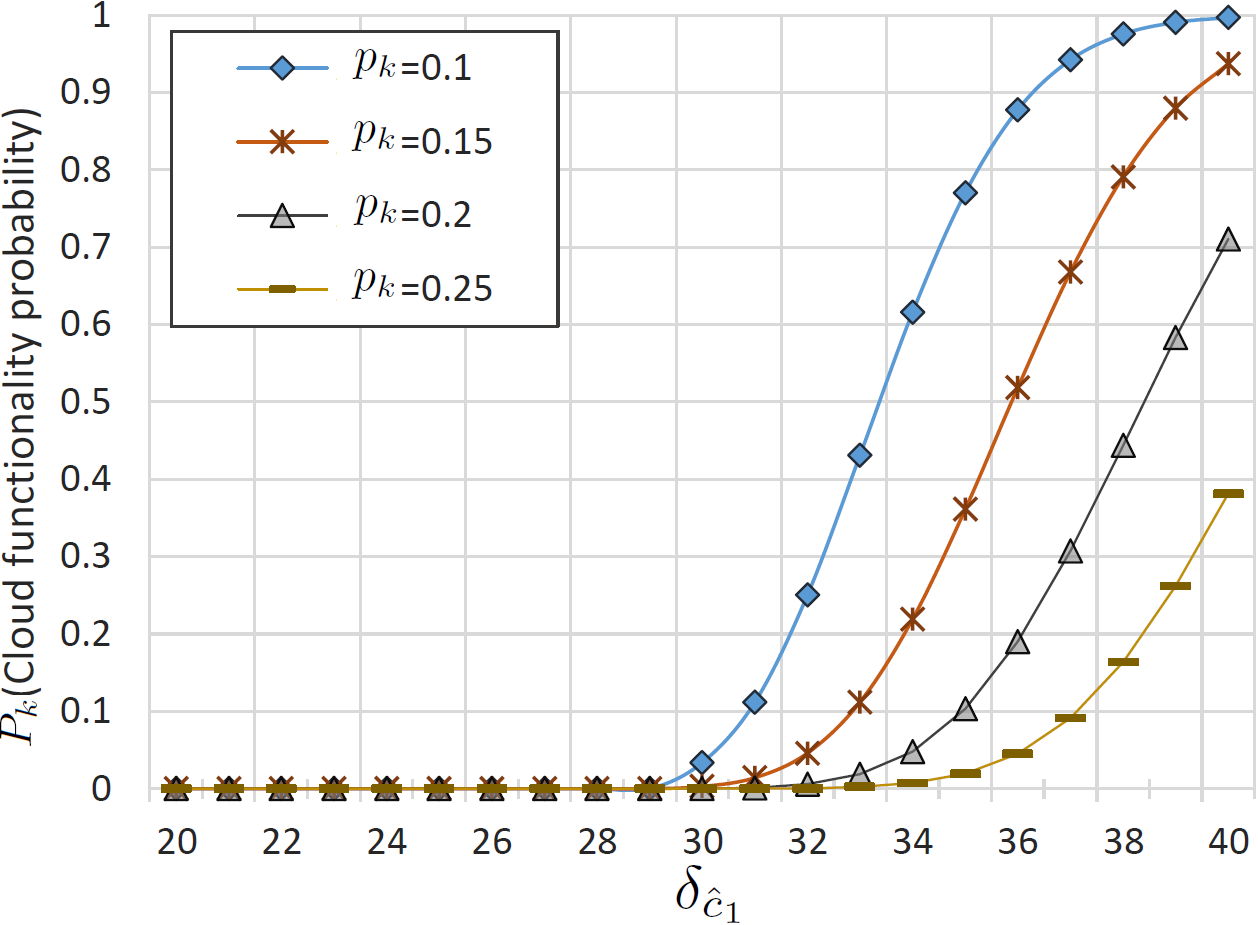}
		\caption{$P_{\hat{c}_1}(\alpha_{\hat{c}_1}=50,\delta_{\hat{c}_1}),~min_{\hat{c}_1}=30$}
		\label{a03}
	\end{subfigure}
	\vspace{3mm}
	
	\caption{Comparing the evaluation results based on various number of defense on cloud-band $\hat{c}_1$ ($20\le\delta_{\hat{c}_1}\le 42 )$ based on various attack success rate $p_k$ values: $0.1, 0.15, 0.20, 0.25$. (a), (b), and (c) show cloud functionality probability ($P_k$) based on various values of $p_k$ and minimum number of maintenance $min_k$ against attack number $\alpha_{\hat{c}_1}=30$. (d), (e), and (f) represent cloud functionality probability ($P_k$) based on various values of $p_k$ and minimum number of maintenance $min_k$ against attack number $\alpha_{\hat{c}_1}=50$.}
	\label{fig:parameters-analysis}
\end{figure*}

\emph{Scenario (\romannumeral 2).} Figure~\ref{c02} compares the results of cloud functionality probability values for a cloud that needs at least 25 running VMs to avoid breakdown. The results are conducted based on 30 attacks and four different attack success rates. The results show that the appropriate number of defense for $P_k$ values of 0.1, 0.15, 0.2, 0.25 are 28, 30, 32, 34 respectively to keep the cloud system functional with the probability of higher $90\%$. It is clear that assigning less than 25 defenses under these circumstances leaves the cloud system under an unreliable condition with a cloud functionality probability of lower $10\%$.

\emph{Scenario (\romannumeral 3).} Figure~\ref{c03} illustrates the results of cloud functionality probability values for a cloud that needs a minimum of 30 running VMs. The results are conducted based on various numbers of defense while 30 attacks launched toward the cloud-based on four different attack success rates. The results show that the appropriate number of defense for $P_k$ values of 0.1, 0.15, 0.2, 0.25 are 34, 36, 38, 40 respectively to keep the cloud system functional with the probability of higher $90\%$. It is clear that assigning less than 30 defenses under these circumstances leaves the cloud system under an unreliable condition with a cloud functionality probability of lower $10\%$.

\emph{Scenario (\romannumeral 4).} Figure~\ref{a01} demonstrates the results of cloud functionality probability values for a cloud that needs a minimum of 20 running VMs. The evaluation is based on 50 attacks toward the cloud with different rates of attack success probability. The number of defense is between 20 and 40. The results show that the appropriate number of defense to keep the cloud functionality more than $90\%$ under different $P_k$ values of 0.1, 0.15, 0.2, 0.25 are 27, 30, 33, 35 respectively. It is clear that assigning less than 20 defenses under these circumstances leaves the cloud system under an unreliable condition with cloud functionality probability of lower $10\%$ while assigning more than 37 defense is not reasonable and affordable based on the defined parameters.

\emph{Scenario (\romannumeral 5).} Figure~\ref{a02} demonstrates the results of cloud functionality probability values for a cloud that needs a minimum of 25 running VMs under 50 number of attacks with different attack success rates. The results show that the appropriate number of defenses to keep the cloud functionality more than $90\%$ in this scenario under different $P_k$ values of 0.1, 0.15, 0.2, 0.25 are 33, 35, 37, 40 respectively. It is clear that assigning less than 25 defenses under these circumstances leaves the cloud system under an unreliable condition with a cloud functionality probability of lower $10\%$.

\emph{Scenario (\romannumeral 6).} Figure~\ref{a03} demonstrates the results of cloud functionality probability values for a cloud that needs a minimum of 30 running VMs under 50 number of attacks with different attack success rates. The results show that the appropriate number of defense to keep the cloud functionality more than $90\%$ in this scenario under different $P_k$ values of 0.1 and 0.15 are 36 and 40 respectively. However, the 40 number of defense for the attack rates of 0.2 and 0.25 leads to lower cloud functionality probability with the values of $70\%$ and $40\%$, respectively.

Based on the results discussed in the above scenarios, we conclude that different attack rates need to be considered and evaluated before denoting the number of defense strategies to the cloud.

\pgfplotsset{
    colormap={parula}{
        rgb=(0.2081,0.1663,0.5292)
        rgb=(0.2116,0.1898,0.5777)
        rgb=(0.2123,0.2138,0.627)
        rgb=(0.2081,0.2386,0.6771)
        rgb=(0.1959,0.2645,0.7279)
        rgb=(0.1707,0.2919,0.7792)
        rgb=(0.1253,0.3242,0.8303)
        rgb=(0.0591,0.3598,0.8683)
        rgb=(0.0117,0.3875,0.882)
        rgb=(0.006,0.4086,0.8828)
        rgb=(0.0165,0.4266,0.8786)
        rgb=(0.0329,0.443,0.872)
        rgb=(0.0498,0.4586,0.8641)
        rgb=(0.0629,0.4737,0.8554)
        rgb=(0.0723,0.4887,0.8467)
        rgb=(0.0779,0.504,0.8384)
        rgb=(0.0793,0.52,0.8312)
        rgb=(0.0749,0.5375,0.8263)
        rgb=(0.0641,0.557,0.824)
        rgb=(0.0488,0.5772,0.8228)
        rgb=(0.0343,0.5966,0.8199)
        rgb=(0.0265,0.6137,0.8135)
        rgb=(0.0239,0.6287,0.8038)
        rgb=(0.0231,0.6418,0.7913)
        rgb=(0.0228,0.6535,0.7768)
        rgb=(0.0267,0.6642,0.7607)
        rgb=(0.0384,0.6743,0.7436)
        rgb=(0.059,0.6838,0.7254)
        rgb=(0.0843,0.6928,0.7062)
        rgb=(0.1133,0.7015,0.6859)
        rgb=(0.1453,0.7098,0.6646)
        rgb=(0.1801,0.7177,0.6424)
        rgb=(0.2178,0.725,0.6193)
        rgb=(0.2586,0.7317,0.5954)
        rgb=(0.3022,0.7376,0.5712)
        rgb=(0.3482,0.7424,0.5473)
        rgb=(0.3953,0.7459,0.5244)
        rgb=(0.442,0.7481,0.5033)
        rgb=(0.4871,0.7491,0.484)
        rgb=(0.53,0.7491,0.4661)
        rgb=(0.5709,0.7485,0.4494)
        rgb=(0.6099,0.7473,0.4337)
        rgb=(0.6473,0.7456,0.4188)
        rgb=(0.6834,0.7435,0.4044)
        rgb=(0.7184,0.7411,0.3905)
        rgb=(0.7525,0.7384,0.3768)
        rgb=(0.7858,0.7356,0.3633)
        rgb=(0.8185,0.7327,0.3498)
        rgb=(0.8507,0.7299,0.336)
        rgb=(0.8824,0.7274,0.3217)
        rgb=(0.9139,0.7258,0.3063)
        rgb=(0.945,0.7261,0.2886)
        rgb=(0.9739,0.7314,0.2666)
        rgb=(0.9938,0.7455,0.2403)
        rgb=(0.999,0.7653,0.2164)
        rgb=(0.9955,0.7861,0.1967)
        rgb=(0.988,0.8066,0.1794)
        rgb=(0.9789,0.8271,0.1633)
        rgb=(0.9697,0.8481,0.1475)
        rgb=(0.9626,0.8705,0.1309)
        rgb=(0.9589,0.8949,0.1132)
        rgb=(0.9598,0.9218,0.0948)
        rgb=(0.9661,0.9514,0.0755)
        rgb=(0.9763,0.9831,0.0538)
    }
}

	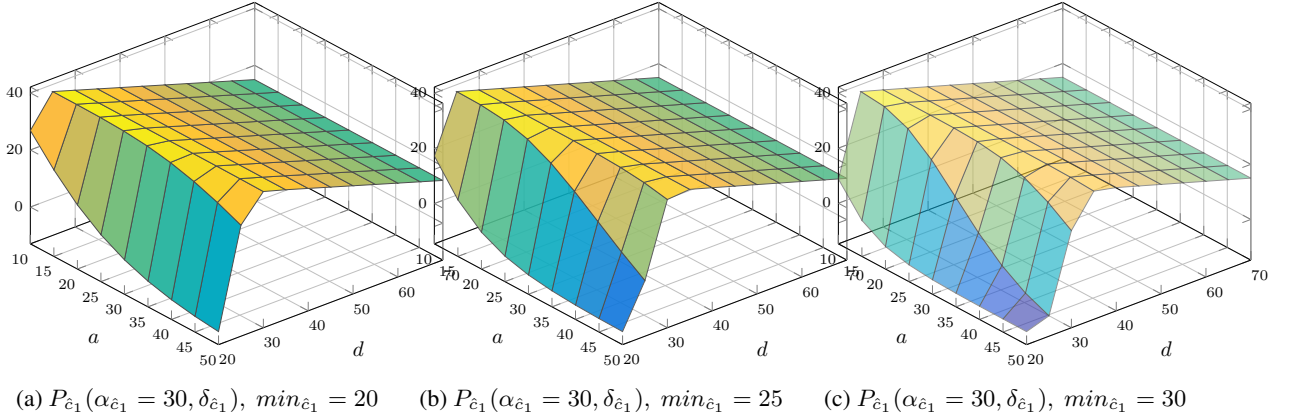
\begin{figure*}[t]
	\centering
	\begin{subfigure}{0.32\textwidth}
\begin{tikzpicture}[scale=0.8]
\begin{axis} [
    view = {50}{40},
    xtick = {10,15,...,50},
    ytick = {20,30,...,70},
    xlabel = $a$, ylabel = $d$,
    ticklabel style = {font = \scriptsize},
	grid,
]
\addplot3 [surf,faceted color=black!70!white] 
	table [
	x=x,
	y=y,
	z=z,
		] {Uk-10-50-pk-0.1.txt};
\end{axis}
\end{tikzpicture}
		\caption{$P_{\hat{c}_1}(\alpha_{\hat{c}_1}=30,\delta_{\hat{c}_1}),~min_{\hat{c}_1}=20$}
		\label{c01}
	\end{subfigure}
	\begin{subfigure}{0.32\textwidth}
\begin{tikzpicture}[scale=0.8]
\begin{axis} [
    view = {50}{40},
    xtick = {10,15,...,50},
    ytick = {20,30,...,70},
    xlabel = $a$, ylabel = $d$,
    ticklabel style = {font = \scriptsize},
	grid
]
\addplot3 [surf,faceted color=black!70!white,opacity=0.9] 
	table [
	x=x,
	y=y,
	z=z,
		] {Uk-10-50-pk-0.15.txt};
\end{axis}
\end{tikzpicture}
		\caption{$P_{\hat{c}_1}(\alpha_{\hat{c}_1}=30,\delta_{\hat{c}_1}),~min_{\hat{c}_1}=25$}
		\label{c02}
	\end{subfigure}
	\begin{subfigure}{0.32\textwidth}
\begin{tikzpicture}[scale=0.8]
\begin{axis} [
    view = {50}{40},
    xtick = {10,15,...,50},
    ytick = {20,30,...,70},
    xlabel = $a$, ylabel = $d$,
    ticklabel style = {font = \scriptsize},
	grid
]
\addplot3 [surf,faceted color=black!70!white,opacity=0.6] 
	table [
	x=x,
	y=y,
	z=z,
		] {Uk-10-50-pk-0.2.txt};
\end{axis}
\end{tikzpicture}
		\caption{$P_{\hat{c}_1}(\alpha_{\hat{c}_1}=30,\delta_{\hat{c}_1}),~min_{\hat{c}_1}=30$}
		\label{c04}
	\end{subfigure}
	\vspace{3mm}
	
	\caption{Comparing the evaluation optimal defender's utility results (y-axis) based on various number of defense and attacks on cloud-band $\hat{c}_1$ ($20\le\delta_{\hat{c}_1}\le 70 )$ based on various attack success rate values: (a) $p_k$: 0.1, (b) $p_k$: 0.15, and (c) $p_k$: 0.2}
	\label{fig:parameters-analysis}
\end{figure*}

\subsection{Discussion and Limitations}
In this paper, we evaluated both attacker's and defender's best strategies against each other while the attacker benefit from an AI-powered botnet attack toward a cloud system. The attacker is equipped with a deep neural network model to conceal the botmaster and bots communication and find and launch DDoS attacks on various critical targets on a cloud model. We summarize the main phases of a new generation of botnet threats:
\begin{itemize}
\item \textit{Neural Network Model:} A neural network model is used to protect accounts of botmasters and conceal the intent of bots, as it is very difficult for adversaries to obtain the accounts in advance even if they get the model and vectors.
\item \textit{Covert Communication}: Then bots start identifying botmaster by its avatar. As the botmaster's avatar is converted to a vector and distributed with bots, bots can pick the botmaster up quickly by comparing distances using the neural network model.
\item \textit{Commands Embeddings}: Target IP addresses are embedded into tweets generated by EDA through hash collision. The Target-embedded tweets are posted by the botmaster. After bots found botmaster, commands can be obtained by calculating hashes of the tweets.
\end{itemize}

However, detection of a new generation of botnet threats is a challenging task due to the characteristics and behaviors of these threats summarized as follows.

\begin{itemize}
\item \textit{Neural network inabilities:} Current proposed neural network models are not able to detect novel threats. The detection of new generation botnets is difficult as there are not enough data about the emerging botnets to train the neural network models. Thus, these models fail an accurate detection or prevention.
\item \textit{Communication detection limitation}: One of the important methods to defend against botnets is to detect the botmaster. The detection of the botnet is performed through the discovery of communication of bots and botmaster by analyzing the behaviors in the communication stages (C\&C). A new generation of botnets utilized concealment techniques for covert communication.
\item \textit{Botnet identification difficulties}: As the novel botnets utilize the DNN model to hide the identity of botmaster and no longer use hardcoding identification methods in the bots, detection of botmaster is very difficult based on current techniques.
\end{itemize}

While it is very difficult to monitor, detect, or prevent this new generation of AI-based botnet threats based on the current defensive methods, one can still benefit from the key benefit of game theory to evaluate the overall impact of such an attack on a system based on both attack and defense to find out the best defensive strategies to mitigate the attack impacts (such as avoiding system failure or breakdown as discussed in Section \ref{sec:GT}). However, the main limitation of game theory in this paper is that the game model is only used to evaluate the best strategies for the defenders to adjust the number of defenses against DDoS attack flows on many targets. However, a more useful defensive method needs to be deployed in addition to an optimal number of defenses to benefit the defensive techniques. Moreover, the attack-defense analysis on the proposed game in this paper assumed that both attacker and defender only conduct one-shot confrontation. However, the model needs to be extended as a dynamic attack-defense game model and treated as a contentious and multi-stage process. Finally, in this study, we only assumed that attacks are launching from OSN to the simulated cloud. However, we didn't simulate the OSN botnet in this study and only utilized theoretical evaluation based on the assumption of different attack rates from botnet to the cloud. We plan to simulate OSN botnet against a realistic cloud testbed in our future work.

Beyond the proposed game theory model to evaluate the best defense strategies against AI-powered botnet threats in this work, we believe it is incumbent on us to discuss other possible countermeasures to provide an insight for future research and direction as follows:

\begin{itemize}
\item \textit{Proactive defensive methods:} Leveraging the key benefit of proactive defensive methods such as Moving Target Defense. MTD provides a new perspective of a defense system by continuously changing the attack surface, which makes it harder for attackers to achieve their goals. This benefits defender based on the following: (i) MTD can ruin the information gathered by botmaster in the botnet reconnaissance phase explained in Section \ref{sec:threats} which botmaster spend time and effort to obtain information about the critical target machines, (ii) as concealing the targets is a time-consuming task, MTD causes the target-embedded comments in OSN to become useless for bots. In this case, bots may spend time extracting the targets and launch attacks on the targets which have already been changed by MTD.
\item \textit{New IDS generation:} New IDS techniques equipped with deep learning utilizing Generative adversarial networks (GANs) can potentially enhance the capability of detecting new attack features and hence increase the detection rate. New IDS generation can also be incorporated and evaluated using the game theory models.
\end{itemize}

\section{Conclusion}\label{sec:conclusion} 
AI-powered botnets leverage DNN techniques to conceal their identity and intentions. While the detection of a new generation of botnet attacks is difficult based on the current defensive methods, the impact of the new generation of botnet attacks toward a target system such as the cloud needs to be evaluated to help the defender to mitigate the risks resulting from these attacks. In this paper, we formulate a sequential game theory model to evaluate both the best defense and attack strategies in terms of the number of attacks and defenses for AI-powered botnet attacks toward a cloud system. We formalized the defender’s optimal strategy based on the number of effective defenses by considering the cost of defense to avoid the cloud system's breakdown while the attacker launches various DDoS attacks toward the critical components of the cloud.

\bibliographystyle{unsrtnat}
\bibliography{IEEEabrv,GT-attack}  

\begin{thebibliography}{41}
\providecommand{\natexlab}[1]{#1}
\providecommand{\url}[1]{\texttt{#1}}
\expandafter\ifx\csname urlstyle\endcsname\relax
  \providecommand{\doi}[1]{doi: #1}\else
  \providecommand{\doi}{doi: \begingroup \urlstyle{rm}\Url}\fi

\bibitem[Hou et~al.(2020)Hou, Li, Cui, Meng, Zhang, Ni, and Tian]{hou2020low}
Jun Hou, Qianmu Li, Shicheng Cui, Shunmei Meng, Sainan Zhang, Zhen Ni, and
  Ye~Tian.
\newblock Low-cohesion differential privacy protection for industrial internet.
\newblock \emph{The Journal of Supercomputing}, 76\penalty0 (11):\penalty0
  8450--8472, 2020.

\bibitem[Brundage et~al.(2018)Brundage, Avin, Clark, Toner, Eckersley,
  Garfinkel, Dafoe, Scharre, Zeitzoff, Filar, et~al.]{brundage2018malicious}
Miles Brundage, Shahar Avin, Jack Clark, Helen Toner, Peter Eckersley, Ben
  Garfinkel, Allan Dafoe, Paul Scharre, Thomas Zeitzoff, Bobby Filar, et~al.
\newblock The malicious use of artificial intelligence: Forecasting,
  prevention, and mitigation.
\newblock \emph{arXiv preprint arXiv:1802.07228}, 2018.

\bibitem[Jang-Jaccard and Nepal(2014)]{jang2014survey}
Julian Jang-Jaccard and Surya Nepal.
\newblock A survey of emerging threats in cybersecurity.
\newblock \emph{Journal of Computer and System Sciences}, 80\penalty0
  (5):\penalty0 973--993, 2014.

\bibitem[Camp et~al.(2019)Camp, Grobler, Jang-Jaccard, Probst, Renaud, and
  Watters]{camp2019measuring}
L~Jean Camp, Marthie Grobler, Julian Jang-Jaccard, Christian Probst, Karen
  Renaud, and Paul Watters.
\newblock Measuring human resilience in the face of the global epidemiology of
  cyber attacks.
\newblock In \emph{Proceedings of the 52nd Hawaii International Conference on
  System Sciences}, 2019.

\bibitem[Stoecklin(2018)]{stoecklin2018deeplocker}
Marc~Ph Stoecklin.
\newblock Deeplocker: How ai can power a stealthy new breed of malware.
\newblock \emph{Security Intelligence, August}, 8, 2018.

\bibitem[Kaloudi and Li(2020)]{kaloudi2020ai}
Nektaria Kaloudi and Jingyue Li.
\newblock The {AI}-based cyber threat landscape: A survey.
\newblock \emph{ACM Computing Surveys (CSUR)}, 53\penalty0 (1):\penalty0 1--34,
  2020.

\bibitem[Wang et~al.(2020)Wang, Liu, Cui, Zhang, Wu, Yin, Liu, Liu, and
  Zhang]{wang2020ai}
Zhi Wang, Chaoge Liu, Xiang Cui, Jialong Zhang, Di~Wu, Jie Yin, Jiaxi Liu, Qixu
  Liu, and Jinli Zhang.
\newblock Ai-powered covert botnet command and control on osns.
\newblock \emph{arXiv preprint arXiv:2009.07707}, 2020.

\bibitem[Singh et~al.(2019)Singh, Singh, and Kaur]{singh2019issues}
Manmeet Singh, Maninder Singh, and Sanmeet Kaur.
\newblock Issues and challenges in dns based botnet detection: A survey.
\newblock \emph{Computers \& Security}, 86:\penalty0 28--52, 2019.

\bibitem[Faghani and Nguyen(2019)]{faghani2019mobile}
Mohammad~R Faghani and Uyen~T Nguyen.
\newblock Mobile botnets meet social networks: design and analysis of a new
  type of botnet.
\newblock \emph{International Journal of Information Security}, 18\penalty0
  (4):\penalty0 423--449, 2019.

\bibitem[Xing et~al.(2021)Xing, Shu, Zhao, Li, and Guo]{xing2021survey}
Ying Xing, Hui Shu, Hao Zhao, Dannong Li, and Li~Guo.
\newblock Survey on botnet detection techniques: Classification, methods, and
  evaluation.
\newblock \emph{Mathematical Problems in Engineering}, 2021, 2021.

\bibitem[Wei et~al.(2021)Wei, Jang-Jaccard, Sabrina, Singh, Xu, and
  Camtepe]{wei2021ae}
Yuanyuan Wei, Julian Jang-Jaccard, Fariza Sabrina, Amardeep Singh, Wen Xu, and
  Seyit Camtepe.
\newblock Ae-mlp: A hybrid deep learning approach for ddos detection and
  classification.
\newblock \emph{IEEE Access}, 2021.

\bibitem[Xu et~al.(2021)Xu, Jang-Jaccard, Singh, Wei, and
  Sabrina]{xu2021improving}
Wen Xu, Julian Jang-Jaccard, Amardeep Singh, Yuanyuan Wei, and Fariza Sabrina.
\newblock Improving performance of autoencoder-based network anomaly detection
  on nsl-kdd dataset.
\newblock \emph{IEEE Access}, 9:\penalty0 140136--140146, 2021.

\bibitem[Zhu et~al.(2020)Zhu, Jang-Jaccard, and Watters]{zhu2020multi}
Jinting Zhu, Julian Jang-Jaccard, and Paul~A Watters.
\newblock Multi-loss siamese neural network with batch normalization layer for
  malware detection.
\newblock \emph{IEEE Access}, 8:\penalty0 171542--171550, 2020.

\bibitem[Zhu et~al.(2021)Zhu, Jang-Jaccard, Singh, Watters, and
  Camtepe]{zhu2021task}
Jinting Zhu, Julian Jang-Jaccard, Amardeep Singh, Paul~A Watters, and Seyit
  Camtepe.
\newblock Task-aware meta learning-based siamese neural network for classifying
  obfuscated malware.
\newblock \emph{arXiv preprint arXiv:2110.13409}, 2021.

\bibitem[Hou et~al.(2019)Hou, Li, Tan, Meng, Zhang, and
  Zhang]{hou2019intrusion}
Jun Hou, Qianmu Li, Rong Tan, Shunmei Meng, Hanrui Zhang, and Sainan Zhang.
\newblock An intrusion tracking watermarking scheme.
\newblock \emph{IEEE Access}, 7:\penalty0 141438--141455, 2019.

\bibitem[Shan and Zhuang(2020)]{shan2020game}
Xiaojun~Gene Shan and Jun Zhuang.
\newblock A game-theoretic approach to modeling attacks and defenses of smart
  grids at three levels.
\newblock \emph{Reliability Engineering \& System Safety}, 195:\penalty0
  106683, 2020.

\bibitem[Manshaei et~al.(2013)Manshaei, Zhu, Alpcan, Ba{\c{s}}ar, and
  Hubaux]{manshaei2013game}
Mohammad~Hossein Manshaei, Quanyan Zhu, Tansu Alpcan, Tamer Ba{\c{s}}ar, and
  Jean-Pierre Hubaux.
\newblock Game theory meets network security and privacy.
\newblock \emph{ACM Computing Surveys (CSUR)}, 45\penalty0 (3):\penalty0 1--39,
  2013.

\bibitem[Attiah et~al.(2018)Attiah, Chatterjee, and Zou]{attiah2018game}
Afraa Attiah, Mainak Chatterjee, and Cliff~C Zou.
\newblock A game theoretic approach to model cyber attack and defense
  strategies.
\newblock In \emph{2018 IEEE International Conference on Communications (ICC)},
  pages 1--7. IEEE, 2018.

\bibitem[Rass et~al.(2017)Rass, Alshawish, Abid, Schauer, Zhu, and
  De~Meer]{rass2017physical}
Stefan Rass, Ali Alshawish, Mohamed~Amine Abid, Stefan Schauer, Quanyan Zhu,
  and Hermann De~Meer.
\newblock Physical intrusion games—optimizing surveillance by simulation and
  game theory.
\newblock \emph{IEEE Access}, 5:\penalty0 8394--8407, 2017.

\bibitem[Lin et~al.(2012)Lin, Liu, and Jing]{lin2012using}
Jingqiang Lin, Peng Liu, and Jiwu Jing.
\newblock Using signaling games to model the multi-step attack-defense
  scenarios on confidentiality.
\newblock In \emph{International Conference on Decision and Game Theory for
  Security}, pages 118--137. Springer, 2012.

\bibitem[Pantic and Husain(2015)]{pantic2015covert}
Nick Pantic and Mohammad~I Husain.
\newblock Covert botnet command and control using twitter.
\newblock In \emph{Proceedings of the 31st annual computer security
  applications conference}, pages 171--180, 2015.

\bibitem[Maim{\'o} et~al.(2018)Maim{\'o}, G{\'o}mez, Clemente, P{\'e}rez, and
  P{\'e}rez]{maimo2018self}
Lorenzo~Fern{\'a}ndez Maim{\'o}, {\'A}ngel Luis~Perales G{\'o}mez, F{\'e}lix
  J~Garc{\'\i}a Clemente, Manuel~Gil P{\'e}rez, and Gregorio~Mart{\'\i}nez
  P{\'e}rez.
\newblock A self-adaptive deep learning-based system for anomaly detection in
  5g networks.
\newblock \emph{IEEE Access}, 6:\penalty0 7700--7712, 2018.

\bibitem[Nguyen et~al.(2020)Nguyen, Ngo, Nguyen, and Le]{nguyen2020psi}
Huy-Trung Nguyen, Quoc-Dung Ngo, Doan-Hieu Nguyen, and Van-Hoang Le.
\newblock Psi-rooted subgraph: A novel feature for iot botnet detection using
  classifier algorithms.
\newblock \emph{ICT Express}, 6\penalty0 (2):\penalty0 128--138, 2020.

\bibitem[Pei et~al.(2020)Pei, Tian, Yu, Wang, and Peng]{pei2020two}
Xinjun Pei, Shengwei Tian, Long Yu, Huanhuan Wang, and Yongfang Peng.
\newblock A two-stream network based on capsule networks and sliced recurrent
  neural networks for dga botnet detection.
\newblock \emph{Journal of Network and Systems Management}, 28\penalty0
  (4):\penalty0 1694--1721, 2020.

\bibitem[Guang et~al.(2016)Guang, TANG, Shuo, SONG, and Yuan]{guang2016using}
KOU Guang, Guang-ming TANG, WANG Shuo, Hai-tao SONG, and BIAN Yuan.
\newblock Using deep learning for detecting botcloud.
\newblock \emph{Journal on Communications}, 37\penalty0 (11):\penalty0 114,
  2016.

\bibitem[Vinayakumar et~al.(2020)Vinayakumar, Alazab, Srinivasan, Pham,
  Padannayil, and Simran]{vinayakumar2020visualized}
R~Vinayakumar, Mamoun Alazab, Sriram Srinivasan, Quoc-Viet Pham, Soman~Kotti
  Padannayil, and K~Simran.
\newblock A visualized botnet detection system based deep learning for the
  internet of things networks of smart cities.
\newblock \emph{IEEE Transactions on Industry Applications}, 56\penalty0
  (4):\penalty0 4436--4456, 2020.

\bibitem[Feng et~al.(2017)Feng, Shuo, and Xiaochuan]{feng2017classification}
Z~Feng, C~Shuo, and W~Xiaochuan.
\newblock Classification for dga-based malicious domain names with deep
  learning architectures.
\newblock In \emph{2017 Second International Conference on Applied Mathematics
  and information technology}, page~5, 2017.

\bibitem[Gaonkar et~al.(2020)Gaonkar, Dessai, Costa, Borkar, Aswale, and
  Shetgaonkar]{gaonkar2020survey}
Shivani Gaonkar, Nandini~Fal Dessai, Jenny Costa, Ashlesha Borkar, Shailendra
  Aswale, and Pratiksha Shetgaonkar.
\newblock A survey on botnet detection techniques.
\newblock In \emph{2020 International Conference on Emerging Trends in
  Information Technology and Engineering (ic-ETITE)}, pages 1--6. IEEE, 2020.

\bibitem[Yin et~al.(2018)Yin, Lv, Zhang, Tian, and Cui]{yin2018study}
Jie Yin, Heyang Lv, Fangjiao Zhang, Zhihong Tian, and Xiang Cui.
\newblock Study on advanced botnet based on publicly available resources.
\newblock In \emph{International Conference on Information and Communications
  Security}, pages 57--74. Springer, 2018.

\bibitem[Chukwudi et~al.(2017)Chukwudi, Udoka, and Charles]{chukwudi2017game}
Amadi~Emmanuuel Chukwudi, Eze Udoka, and Ikerionwu Charles.
\newblock Game theory basics and its application in cyber security.
\newblock \emph{Advances in Wireless Communications and Networks}, 3\penalty0
  (4):\penalty0 45--49, 2017.

\bibitem[Asadi(2021)]{asadi2021detecting}
Mehdi Asadi.
\newblock Detecting iot botnets based on the combination of cooperative game
  theory with deep and machine learning approaches.
\newblock \emph{Journal of Ambient Intelligence and Humanized Computing}, pages
  1--15, 2021.

\bibitem[Wang et~al.(2016)Wang, Ma, Zhang, Ji, Lu, and Hei]{wang2016dynamic}
Yichuan Wang, Jianfeng Ma, Liumei Zhang, Wenjiang Ji, Di~Lu, and Xinhong Hei.
\newblock Dynamic game model of botnet ddos attack and defense.
\newblock \emph{Security and Communication Networks}, 9\penalty0 (16):\penalty0
  3127--3140, 2016.

\bibitem[Chauhan(2018)]{chauhan2018practical}
Ajay~Singh Chauhan.
\newblock \emph{Practical Network Scanning: Capture network vulnerabilities
  using standard tools such as Nmap and Nessus}.
\newblock Packt Publishing Ltd, 2018.

\bibitem[Ghosh et~al.(2021)Ghosh, Kumar, Bhatia, and
  Vishwakarma]{ghosh2021using}
Indraneel Ghosh, Subham Kumar, Ashutosh Bhatia, and Deepak~Kumar Vishwakarma.
\newblock Using auxiliary inputs in deep learning models for detecting
  dga-based domain names.
\newblock In \emph{2021 International Conference on Information Networking
  (ICOIN)}, pages 391--396. IEEE, 2021.

\bibitem[Wei and Zou(2019)]{WeiZ19}
Jason~W. Wei and Kai Zou.
\newblock {EDA:} easy data augmentation techniques for boosting performance on
  text classification tasks.
\newblock In Kentaro Inui, Jing Jiang, Vincent Ng, and Xiaojun Wan, editors,
  \emph{Proceedings of the 2019 Conference on Empirical Methods in Natural
  Language Processing and the 9th International Joint Conference on Natural
  Language Processing, {EMNLP-IJCNLP} 2019, Hong Kong, China, November 3-7,
  2019}, pages 6381--6387. Association for Computational Linguistics, 2019.
\newblock \doi{10.18653/v1/D19-1670}.
\newblock URL \url{https://doi.org/10.18653/v1/D19-1670}.

\bibitem[Kolias et~al.(2017)Kolias, Kambourakis, Stavrou, and
  Voas]{kolias2017ddos}
Constantinos Kolias, Georgios Kambourakis, Angelos Stavrou, and Jeffrey Voas.
\newblock Ddos in the iot: Mirai and other botnets.
\newblock \emph{Computer}, 50\penalty0 (7):\penalty0 80--84, 2017.

\bibitem[Sanjab and Saad(2016)]{sanjab2016bounded}
Anibal Sanjab and Walid Saad.
\newblock On bounded rationality in cyber-physical systems security:
  Game-theoretic analysis with application to smart grid protection.
\newblock In \emph{2016 Joint Workshop on Cyber-Physical Security and
  Resilience in Smart Grids (CPSR-SG)}, pages 1--6. IEEE, 2016.

\bibitem[Alavizadeh et~al.(2021)Alavizadeh, Hong, Kim, and
  Jang-Jaccard]{alavizadeh2021evaluating}
Hooman Alavizadeh, Jin~B Hong, Dong~Seong Kim, and Julian Jang-Jaccard.
\newblock Evaluating the effectiveness of shuffle and redundancy mtd techniques
  in the cloud.
\newblock \emph{Computers \& Security}, 102:\penalty0 102091, 2021.

\bibitem[Alavizadeh et~al.(2019)Alavizadeh, Alavizadeh, Kim, Jang-Jaccard, and
  Torshiz]{alavizadeh2019automated}
Hooman Alavizadeh, Hootan Alavizadeh, Dong~Seong Kim, Julian Jang-Jaccard, and
  Masood~Niazi Torshiz.
\newblock An automated security analysis framework and implementation for mtd
  techniques on cloud.
\newblock In \emph{International Conference on Information Security and
  Cryptology}, pages 150--164. Springer, 2019.

\bibitem[Alavizadeh et~al.(2018)Alavizadeh, Jang-Jaccard, and
  Kim]{alavizadeh2018evaluation}
Hooman Alavizadeh, Julian Jang-Jaccard, and Dong~Seong Kim.
\newblock Evaluation for combination of shuffle and diversity on moving target
  defense strategy for cloud computing.
\newblock In \emph{2018 17th IEEE International Conference On Trust, Security
  And Privacy In Computing And Communications/12th IEEE International
  Conference On Big Data Science And Engineering (TrustCom/BigDataSE)}, pages
  573--578. IEEE, 2018.

\bibitem[Alavizadeh et~al.(2017)Alavizadeh, Kim, Hong, and
  Jang-Jaccard]{alavizadeh2017effective}
Hooman Alavizadeh, Dong~Seong Kim, Jin~B Hong, and Julian Jang-Jaccard.
\newblock Effective security analysis for combinations of mtd techniques on
  cloud computing (short paper).
\newblock In \emph{International Conference on Information Security Practice
  and Experience}, pages 539--548. Springer, 2017.

\end{thebibliography}






\end{document}